\documentclass[11pt,twoside,a4paper]{article}

\RequirePackage{fix-cm}
\usepackage{graphicx}
\usepackage{subcaption}
\usepackage{amssymb}
\usepackage{setspace}
\usepackage{booktabs}

\usepackage{fancyhdr}
\newcommand{\eref}[1]{(\ref{#1})}
\newcommand{\vect}[1]{\mathbf{#1}}
\newcommand{\address}[3]{
  \noindent
	\begin{minipage}[t]{0.75\textwidth}
	{\begin{flushleft}\footnotesize
  \par\noindent \textit{#1}
	\par\noindent {#2}
	\par\noindent\texttt{#3}
	\end{flushleft}}
	\end{minipage}
	\vspace{\baselineskip}
}
\newcommand{\keywords}[1]{\par\noindent\textbf{Keywords:} #1}
\newcommand{\affiliations}{\par\vspace{12pt}\noindent\textbf{Authors' affiliations}\par}

\newtheorem{lem}{Lemma}[section]
\newtheorem{defn}{Definition}[section]

\newtheorem{rem}{Remark}[section]

\newtheorem{asm}[lem]{Assumption}

\title{Inversion of a SIR-based model: a critical analysis about the application to COVID-19 epidemic}
\author{M. Giudici$^\diamond$, A. Comunian$^\diamond$, R. Gaburro$^\star$\\
$^\diamond$Universit\`a degli Studi di Milano, Italy\\
$^\star$University of Limerick, Ireland}
\date{\today}

\begin{document}

\maketitle

\pagestyle{fancy}

\begin{abstract}\noindent
Calibration of a SIR (Susceptibles-Infected-Recovered) model with official international data for the COVID-19 pandemics provides a good example of the difficulties inherent the solution of inverse problems. Inverse modeling is set up in a framework of discrete inverse problems, which explicitly considers the role and the relevance of data. Together with a physical vision of the model, the present work addresses numerically the issue of parameters calibration in SIR models, it discusses the uncertainties in the data provided by international authorities, how they influence the reliability of calibrated model parameters and, ultimately, of model predictions.
\end{abstract}


\section{Introduction}\label{intro}

Epidemic modeling is usually performed with compartmental models, often called SIR (Susceptibles-Infected-Recovered) models, which are claimed to go back to the work by Ronald Ross and Hilda P.~Hudson more than one century ago \cite{Ross1916, RossHudson1917} and, ten years later, to the work of Anderson Gray McKendrick and William Ogilvy Kermack \cite{kermack1927contribution, kermack1932contribution}. This class of models shares several characteristics with models of population dynamics and with conceptual lumped models in hydrology. These models simulate the temporal evolution of some compartments of the population, which is normally subdivided among \textit{Susceptibles} (i.e., those individuals who have not yet been affected by the virus and which could be subject to infection), \textit{Infected} (i.e., those individuals who have been infected by the virus) and \textit{Recovered} (i.e., those individuals who have recovered, after having been infected). For this reason, these models are usually referred to as SIR models. They are based on phenomenological laws to describe the transfer of individuals from one class to another.

These models have found wide application both in life sciences, mostly in epidemiology, and in the field of economic, political and social sciences, e.g., in the context of addressing the costs of policies designed to block epidemics and the diffusion of viruses and in the realm of optimal control to assess the political measures which guarantee the best equilibrium between reduction of the epidemic spread and harmful secondary socio-economical impacts \cite{Reluga2010, FENICHEL2013, NBERw26882}. Several extremely interesting papers have been devoted to the mathematical properties of SIR models, often by applying the theory of dynamical systems; see \cite{Capasso1978, capasso1983, BerTak1995, Hethcote2000, KAMO2002, GREENMAN2004, Korobeinikov2006, WANG2007, PENG2013, MCCLUSKEY2010} among many others. However, previous works on SIR models have, to the best of our knowledge, seldom addressed the calibration of SIR models with real data, i.e., the issue of a proper fitting of epidemiological data with model outcomes. Some examples refer to applications to dengue transmission \cite{PANDEY2013}, H5N1 avian influenza \cite{BetRib}, HIV epidemic \cite{Joshi} and Severe Acute Respiratory Syndrome (SARS) \cite{Ng}.

In this paper, we address the problem of calibrating the epidemiological parameters of a SIR model describing the evolution in time of the current COVID-19 pandemic. This is achieved by solving numerically the underlying inverse problem via the minimization of an objective function that measures the discrepancy between our simulated solutions to the discretised SIR model and official data on COVID-19.

Model calibration is a common problem in geophysical and environmental modeling. The present paper follows the general framework introduced in \cite{giudici2019conceptual} to handle discrete inverse problems for model calibration and analyses the role of data following the discussion in \cite{Giudici2001}. The continuum SIR model considered here is discretised via a forward-time finite-differences scheme which is implemented in a specifically designed code, developed using the Python programming language, to provide at each discrete time $n\in\mathbb{Z}$ a vector state of the discretised SIR system, which is, in turn, matched against real data in order to calibrate the parameters  of the system via a minimization problem (the inverse problem). The results presented in this paper consider the application of the model to a given nation, Italy in this instance, i.e., the population of the whole nation is considered, without any further subdivision in provinces, regions or states.

The wide number of data collected during the COVID-19 pandemic due to the diffusion of the SARS-CoV-2 virus (also called ``coronavirus'') provides an exceptional basis to test calibration of SIR models via the solution of an inverse problem. It is well known that inverse problems are ill-posed, due to the lack of uniqueness and stability of these problems. Non-uniqueness will be considered in this paper by application of different algorithms for the minimization of the misfit between reference observed quantities and model predictions. The other relevant topic for ill-posedness is the lack of stability, i.e., the lack of continuous dependence of the parameters to be identified on the data, so that small errors in the data can lead to large discrepancies in the parameters one is trying to identify via the inverse problem. We do not provide a full review here on these topics, but we mention \cite{TA} for a general-purpose description, \cite{Alessandrini1988,Alessandrini2007} for a deep discussion on the instability issue in the context of the so-called inverse conductivity problem and \cite{DGLN} for recent results about optical tomography.

The objectives of this paper are to fix some concepts about SIR models and their calibration and to discuss the relevance of data for reliability of model outcomes in the context of inverse problems. The paper is designed to advance the current knowledge about the functioning, potentialities and limitations of epidemic models. It also highlights certain similarities among geophysical, environmental and epidemic modeling, therefore providing further insights in epidemic model calibration. On the other hand, this work does not aim to provide forecasts of the pandemic evolution at this stage. It is in the authors' opinion that the quality of the data that are currently available does not allow to perform reliable forecast and model outcomes should be used with high prudence. It will be material of future work to further develop our SIR model and to address the issue of providing forecasts of the epidemics, when the data will be better understood.

The paper is organized as follows. Section \ref{sec:methods and materials} contains the description of the SIR model in both the continuous and the discrete case (subsection \ref{sec:Model}) together with a precise formulation of the inverse problem addressed in this paper in the discrete setting (subsection \ref{sec:InverseProblem}).
In particular, inverse modeling, i.e., model calibration, is set up and discussed computationally within the framework proposed by \cite{giudici2019conceptual}. The results obtained by applying our SIR model to the COVID-19 pandemic are shown in section \ref{sec:results}. Section \ref{sec:discussion} is devoted to a discussion about the main assumptions on which the SIR model is based; it also contains some remarks about model calibration and data uncertainty. The concluding section contains a final discussion about various possible future developments of this work.

\section{Methods and materials}\label{sec:methods and materials}

\subsection{The continuous and the discrete models}\label{sec:Model}

We start by defining the objects involved in the continuous SIR model considered in this paper. 

\begin{defn}\label{definizione categorie}
We denote by $S(t)$, $I(t)$, $R(t)$ and $D(t)$ the number of \textit{susceptible}, \textit{infected}, \textit{recovered} and \textit{deceased individuals} of the population under study at time $t$, respectively, for $t$ varying in some interval $\cal{I}\subset\mathbb{R}$. Here $D$ includes only those individuals who died while being infected, whereas the total population, at time $t$, is given by $P(t)=S(t)+I(t)+R(t)$.
\end{defn}

\begin{defn}\label{definizione parametri}
We denote by $\beta$ and $\delta$ the \textit{birth} and \textit{death rate}, respectively, under normal conditions, i.e., without considering deaths caused by the epidemic. We also denote by $\gamma$, $\rho$ and $\phi$ the \textit{infection}, 
\textit{recovery} and \textit{fatality rate}, respectively. The dimension of these coefficients is $[\mathrm{time}^{-1}]$. 
\end{defn}

Note that $\beta$ and $\delta$ in definition \ref{definizione parametri} are rarely considered in epidemic modeling, as the time variation of $P$ due to the normal evolution of the population is either negligible or smoother than its variation due to the presence of an epidemic. This is due to the fact that typical values of $\beta$ and $\delta$ are smaller than the ones of $\gamma$, $\rho$ and $\phi$ by one or more orders of magnitude, as shown in subsection \ref{sec:ModelCalibration}. We keep birth and death rates in the model, in order to facilitate a thorough discussion of the assumptions behind this model, which is given in section \ref{sec:discussion}. We make the following assumptions.
\begin{asm}\label{assumption parametri}
The coefficients $\beta$, $\delta$, $\gamma$, $\rho$ and $\phi$ are assumed to be constant.
\end{asm}

\begin{asm}
The number of contacts of each infected person per unit time does not vary among the infected population and it is assumed to be constant in time. Moreover the fraction of such contacts who are susceptible to the infection is given by $S/P$, whereas $(I+R)/P$ is the fraction of those persons who cannot be infected, as it is also assumed that recovered people are immunized. 
\end{asm}

The following equations, based on the seminal papers \cite{Ross1916, RossHudson1917, kermack1927contribution, kermack1932contribution}, are used to describe the time evolution of $S$, $I$, $D$ and $R$:
\begin{eqnarray}
	  \displaystyle \frac{\mathrm{d}S}{\mathrm{d}t} &=& \beta S - \gamma \displaystyle \frac{IS}{P} - \delta S, 
		\label{eq 1}\\[6pt]
	  \displaystyle \frac{\mathrm{d}I}{\mathrm{d}t} &=& \beta I  + \gamma \displaystyle \frac{IS}{P} - \rho I - \phi I - \delta I,
		\label{eq 2}\\[6pt]
	  \displaystyle \frac{\mathrm{d}D}{\mathrm{d}t} &=& \phi I,
		\label{eq 3}\\[6pt]
	  \displaystyle \frac{\mathrm{d}R}{\mathrm{d}t} &=& \beta R + \rho I - \delta R \label{eq 4}
\end{eqnarray}
together with the initial conditions $S(t_\mathrm{ini})=P_\mathrm{ini} - 1$, $I(t_\mathrm{ini})=1$, $R(t_\mathrm{ini})=0$ and $D(t_\mathrm{ini})=0$, where $t_\mathrm{ini}\in \cal{I} \subset\mathbb{R}$ is the time at which the first individual is infected and $P_\mathrm{ini}$ is the population at $t_\mathrm{ini}$. 
Notice that from equations \eref{eq 1} to \eref{eq 4} one can easily deduce
\begin{equation}\label{eq:P}
	\frac{\mathrm{d}P}{\mathrm{d}t} = \beta P - \delta P - \phi I
\end{equation}
and if we couple \eref{eq:P} with \eref{eq 2}
\begin{equation}\label{system P, I}
\left\{ \begin{array}{ll} 
\displaystyle \frac{\mathrm{d}P}{\mathrm{d}t} = (\beta - \delta) P - \phi I, &\qquad \textnormal{in}\quad \cal{I},\\[6pt]
\displaystyle \frac{\mathrm{d}I}{\mathrm{d}t} = - \alpha I + \gamma \displaystyle \frac{IS}{P} ,&\qquad \textnormal{in}\quad \cal{I},\\[6pt]
P(t_\mathrm{ini})=P_\mathrm{ini},& \\
I(t_\mathrm{ini})=1,&
\end{array} \right.
\end{equation}
where $\alpha = \phi + \rho - \beta + \delta$.

We can approximate \eref{system P, I} to the simple system of autonomous linear ordinary differential equations
\begin{equation}\label{system P, I autonomous}
\left\{ \begin{array}{ll} 
\displaystyle \frac{\mathrm{d}P}{\mathrm{d}t} = (\beta - \delta) P, &\qquad \textnormal{in}\quad (t_\mathrm{ini}
, t_\mathrm{ini} +h),\\[6pt]
\displaystyle \frac{\mathrm{d}I}{\mathrm{d}t} = (\gamma - \alpha) I
,&\qquad \textnormal{in}\quad (t_\mathrm{ini}
, t_\mathrm{ini} + h),\\[6pt]
P(t_\mathrm{ini})=P_\mathrm{ini}, & \\
I(t_\mathrm{ini})=1, &
\end{array} \right.
\end{equation}
for some $h$, $0<h\ll 1$
. This rough approximation is justified by thinking that, for $h$ small enough, $I(t)\ll P_\mathrm{ini}\simeq S(t)$ and therefore $IS/P\simeq I$ in \eref{system P, I}.

The system
\begin{equation}\label{eq:P(t<0)}
\left\{ \begin{array}{ll} 
\displaystyle \frac{\mathrm{d}P}{\mathrm{d}t} = (\beta - \delta) P, &\qquad \textnormal{in}\quad (t_\mathrm{ini}
, t_\mathrm{ini} + h),\\
P(t_\mathrm{ini})=P_\mathrm{ini}, & 
\end{array} \right.
\end{equation}
describes the population evolution taking into account demographic aspects only, i.e., in absence of the perturbation caused by epidemics and by assuming that the birth and death rate are constant, whereas
\begin{equation}\label{eq:I(t<0)}
\left\{ \begin{array}{ll} 
\displaystyle \frac{\mathrm{d}I}{\mathrm{d}t} = (\gamma - \alpha) I ,&\qquad \textnormal{in}\quad 
(t_\mathrm{ini}, t_\mathrm{ini} + h),\\
I(t_\mathrm{ini})=
1 & 
\end{array} \right.
\end{equation}
describes the time evolution of the number of infected cases during a short time 
after the beginning of the infection at time $t=t_\mathrm{ini}$. The solution to \eref{eq:I(t<0)}, $I(t)\simeq \exp\left[ (\gamma - \alpha) \cdot \left(t -t_\mathrm{ini} \right)\right]$ and, for $h$ small enough, its linear approximation near $t_\mathrm{ini}$, $I(t)\simeq 1 + \left( \gamma - \alpha \right) \cdot \left(t - t_\mathrm{ini} \right)$, gives  a first rough explanation about why, during the first phases of the epidemics, i.e., for $t\simeq t_\mathrm{ini}$, the number of infected individuals, $I(t)$, seems to grow linearly. This fact motivates the difficulties in the design of an efficient early-warning system. In fact, once $I(t)$ increases to a significant level to be detected, the exponential growth had already kicked in and the containment measures can be effective only if quite drastic.


The discrete model is a simple forward-time finite-differences discretization of equations \eref{eq 1} to \eref{eq 4}. For $n\in\mathbb{Z}$, we denote the discrete time steps, at a uniform spacing $\Delta t$, by $t_n = n \Delta t$. The following definition is useful for the discrete model.

\begin{defn}\label{definizione categorie discrete}
We denote by $S_n$, $I_n$, $R_n$ and $D_n$ the number of \textit{susceptible}, \textit{infected}, \textit{recovered} and \textit{deceased individuals} of the population under study at time $t_n$, respectively, for $n=n_\mathrm{ini},\ldots,n_\mathrm{ini}+N^\mathrm{(mod)}-1$, where $n_\mathrm{ini}$ is such that $t_\mathrm{ini} = n_\mathrm{ini}\Delta t$ and $N^\mathrm{(mod)}$ is the number of modeled time steps. The total population at time $t_n$ is given by $P_n = S_n + I_n + R_n$.
\end{defn}

Then the resulting algebraic iterative equations are of the form
\begin{equation}\label{eq:DiscreteModel}
  \left\{
	\begin{array}{rcl}
		S_{n+1} &=& \left[1 + \left( \beta - \gamma \displaystyle \frac{I_n}{P_n} - \delta \right) \Delta t\right] S_n,
		\vspace{6pt}\\
	  I_{n+1} &=& \left[1 + \left( \beta + \gamma \displaystyle \frac{S_n}{P_n} - \rho - \phi -\delta \right) \Delta t\right] I_n,
	  \vspace{6pt}\\
		D_{n+1} &=& D_n + \phi I_n \Delta t,
	  \vspace{6pt}\\
		R_{n+1} &=& \left[1 + (\beta - \delta)\Delta t\right] R_n + \rho I_n\Delta t,
  \end{array}
	\right.
\end{equation}
for  $n=n_\mathrm{ini},\ldots,n_\mathrm{ini}+N^\mathrm{(mod)}-1$, with initial conditions
\begin{equation}\label{eq:DiscreteInitialConditions}
S_{n_\mathrm{ini}} = P_\mathrm{ini} - 1,\qquad I_{n_\mathrm{ini}} = 1,\qquad D_{n_\mathrm{ini}} = R_{n_\mathrm{ini}} = 0
\end{equation}
and the discrete counterpart of \eref{eq:P} is
\begin{equation}\label{eq:DiscreteModelP}
		P_{n+1} = \left[1 + (\beta - \delta) \Delta t\right] P_n - \phi I_n \Delta t.
\end{equation}

Here the time spacing $\Delta t = 1\,\mathrm{day}$, in agreement with the sampling of the available data set on COVID-19 pandemic (see section \ref{sec:Data}). Equations \eref{eq:DiscreteModel} are implemented in a specifically designed code, developed using the Python programming language. The choice $n\in\mathbb{Z}$ allows to simplify the notation adopted in the formulation of the inverse problem in section \ref{sec:InverseProblem}. It is important to notice that $n=0$, i.e., $t_0 = 0$, represents the first day for which epidemic data are available and in general it does not coincide with $n=n_\mathrm{ini}$, which corresponds to $t_\mathrm{ini}$, the day when the first person was infected in a given nation, according to our model. We will call $t_0=0$ ($n=0$) and $t_\mathrm{ini}$ ($n=n_\mathrm{ini}$) the \textit{monitoring initial time} and the \textit{modelling initial time}, respectively. Accordingly, we will also call $P_\mathrm{ini} = P(t_\mathrm{ini})$ the \textit{model initial population}.

\subsection{The inverse problem: model calibration}\label{sec:InverseProblem}

As stated in the introduction, the inverse problem addressed here is defined in the discrete setting by making use of the conceptual framework and the notation of \cite{giudici2019conceptual}. The numerical task in treating the inverse problem consists in solving iteratively \eref{eq:DiscreteModel} and matching such solutions with the data collected within a certain time-frame $[t_\mathrm{min}, t_\mathrm{max})$. Such (discrete) time-varying vector-solutions $s_n$ are collected in an array $\vect{s}$, called the \textit{state of the system}
\begin{equation}\label{eq:statearray}
  \begin{array}{rr}
	\vect{s} = & \Big\{ s_n = (s^{(1)}_n, s^{(2)}_{n}, s^{(3)}_{n}, s^{(4)}_{n})\in\mathbb{R}^4 \: |\hfill\\
	& s^{(1)}_{n} = S_n,\: s^{(2)}_{n} = I_n,\: s^{(3)}_{n} = R_n,\: s^{(4)}_{n} = D_n,\\
	& \hfill n = n_\mathrm{ini},\ldots,n_\mathrm{ini}+N^\mathrm{(mod)}-1\Big\},
	\end{array}
\end{equation}
where $N^\mathrm{(mod)}$ and $n=n_\mathrm{ini}$ have been introduced in definition \ref{definizione categorie discrete}.

$\vect{s}$ is the model outcome used to forecast the number of infected, recovered and dead individuals. To this end, we also introduce the \textit {model forecast}, an array $\vect{y}$ defined by
\begin{equation}\label{eq:y}
  \begin{array}{r}
	\vect{y} = \Big\{ y_n= (y_n^{(1)}, y_n^{(2)}, y_n^{(3)})\in\mathbb{R}^3 \: | \:  y_n^{(1)}= I_n,\: y_n^{(2)} = R_n,\: y_n^{(3)}= D_n,\\
	n=n_\mathrm{min},\ldots,n_\mathrm{max}-1\Big\},
  \end{array}
\end{equation}
for some $n_\mathrm{min}$, $n_\mathrm{max}$, with $n_\mathrm{ini}\leq n_\mathrm{min}<n_\mathrm{max}\leq n_\mathrm{ini} + N^\mathrm{(mod)}$.
The available data are collected in an array $\vect{d}$. In the specific case considered here, the subset of \textit{data} denoted by $\vect{d}^\prime \subset \vect{d}$ includes the cumulative number of the confirmed infected cases, together with the number of the recovered and dead persons, released by health official organizations
\begin{equation}\label{eq:data}
	\begin{array}{rr}
	\vect{d}^{\prime} = & \Big\{ d_n^{\prime}= (d_n^{\prime (1)}, d_n^{\prime (2)}, d_n^{\prime (3)})\in\mathbb{R}^3 \: |\hfill\\
	& d_n^{\prime (1)}= C^\mathrm{(ref)}_n,\: d_n^{\prime (2)} = R^\mathrm{(ref)}_n,\: d_n^{\prime (3)}= D^\mathrm{(ref)}_n,\\
	& \hfill n=0,\ldots,N^\mathrm{(ref)}-1\Big\},
	\end{array}
\end{equation}
where $N^\mathrm{(ref)}$ is the number of data time steps, i.e., the number of time steps for which data are available. Notice that $C^\mathrm{(ref)}_n$ is the cumulative number of confirmed infected cases, so that the number of infected cases at a given time $n$ is given by
\begin{equation}\label{C}
I^\mathrm{(ref)}_n  = C^\mathrm{(ref)}_n - R^\mathrm{(ref)}_n - D^\mathrm{(ref)}_n.
\end{equation}
$\vect{d}$ can include also other data, e.g., demographic data used to infer the values of some model parameters ($\beta$ and $\delta$).
$n=0$ represents the so-called \textit{monitoring initial time} introduced in \ref{sec:Model}, which corresponds to the first day for which epidemic data $\vect{d}^\prime$ are available; recall that, in general, it does not coincide with the day $n=n_\mathrm{ini}$ when the first person was infected in a given country. 

Model calibration requires that the model forecast be close to a \textit{calibration target}, an array $\vect{t}$ that collects the values which should be attained by the model forecast, if the model were physically ``correct'' and the model parameters were ``optimal''. In this specific case $\vect{t}$ is defined by
\begin{equation}\label{eq:t}
	\begin{array}{rr}
	  \vect{t} = & \Big\{ t_n= (t_n^{(1)}, t_n^{(2)}, t_n^{(3)})\in\mathbb{R}^3 \: |\hfill\\
	  & t_n^{(1)}= I^\mathrm{(ref)}_n,\: t_n^{(2)} = R^\mathrm{(ref)}_n,\: t_n^{(3)}= D^\mathrm{(ref)}_n,\\
	  & \hfill n=n_\mathrm{min} \ldots, n_\mathrm{max}-1\Big\},
	\end{array}
\end{equation}
where $I^\mathrm{(ref)}_n$ is given by \eref{C} and  $n_\mathrm{min}$, $n_\mathrm{max}$ are such that $0\leq n_\mathrm{min}<n_\mathrm{max}\leq N^\mathrm{(ref)}$.
The \textit{model parameters} are placed in an array $\vect{p}$:
\begin{equation}\label{eq:parameterarray}
	\vect{p} = \left(\beta, \delta, \Delta t, \rho, \phi, \gamma, n_\mathrm{ini}, P_\mathrm{ini}\right) \in {\cal P} \subset {\mathbb{R}_+}^6\times\mathbb{Z}\times(\mathbb{N}\setminus\{0\}),
\end{equation}
where $\mathbb{R}_{+} = (0,+\infty)$ and we recall that $\Delta t = 1\,\mathrm{day}$ and $P_\mathrm{ini}$ is the \textit{model initial population} introduced in section \ref{sec:Model}.

If we summarize the algebraic equations in the discrete model \eref{eq:DiscreteModel} together with the initial conditions \eref{eq:DiscreteInitialConditions} with
\begin{equation}\label{eq:DiscreteModelSynthetic}
	\vect{f}(\vect{p},\vect{s}) = 0,
\end{equation}
the \textit{forward problem} can be stated as: \textit{given} $\vect{p}$, \textit{find the unique state} $\vect{s}=\vect{g}(\vect{p})$ that solves \eref{eq:DiscreteModelSynthetic}. In other words, given the parameters $\vect{p}$, the solution to the forward problem will give the state of the system, $\vect{s}$.  In order to introduce the corresponding \textit{inverse problem}, it is convenient to write $\vect{p}$ as
\begin{equation}\label{p sub}
\vect{p}=\left(\vect{p}^\mathrm{(fix)}, \vect{p}^\mathrm{(cal)}\right), 
\end{equation}
where
\begin{equation}\label{eq:pfix e pcal}
  \vect{p}^\mathrm{(fix)} = \left( \beta, \delta, \Delta t \right),\quad \vect{p}^\mathrm{(cal)} = \left( \rho, \phi, \gamma, n_\mathrm{ini}, P_\mathrm{ini} \right).
\end{equation}
$\vect{p}^\mathrm{(fix)}$ and $\vect{p}^\mathrm{(cal)}$ include the model parameters, whose values are fixed before the simulation and the model parameters whose values are obtained from the solution of the underlying inverse problem, which is yet to be stated, respectively.
\begin{rem}
Some remarks on $\vect{p}, \vect{y}$ and $\vect{t}$ are in order.
\begin{enumerate}
  \item The array of fixed parameters is a function of $\vect{d}$: $\vect{p}^\mathrm{(fix)} = \vect{p}^\mathrm{(fix)}(\vect{d})$;
	\item The model forecast $\vect{y}$ is a function of $\vect{s}$, $\vect{p}$ and $\vect{d}$: $\vect{y}$ = $\vect{y}\left(\vect{d},\vect{s},\vect{p}\right)$;
	\item $\vect{t}$ may depend on $\vect{d}$ and $\vect{p}^\mathrm{(fix)}$, but must be independent of $\vect{p}^\mathrm{(cal)}$: $\vect{t}=\vect{t}\left(\vect{d},\vect{p}^\mathrm{(fix)}\right)$.
\end{enumerate}
\end{rem}
The misfit between model predictions and the target values is computed by means of the following objective function:
\begin{equation}\label{eq:ObjectiveFunction}
\mathsf{O}_{\vect{y},\vect{t}} \left( \vect{p}^\mathrm{(cal)} \right) = \sum_{i=1}^3 \mathsf{O}^{(i)}_{\vect{y},\vect{t}} \left( \vect{p}^\mathrm{(cal)}\right) 
\end{equation}
where $\mathsf{O}^{(i)}_{\vect{y},\vect{t}} \left( \vect{p}^\mathrm{(cal)}\right)$ is defined by
\begin{equation}\label{eq:ObjectiveFunctionExplicit}
	\mathsf{O}^{(i)}_{\vect{y},\vect{t}} \left( \vect{p}^\mathrm{(cal)}\right) = \left\{ \frac{1}{n_\mathrm{max}-n_\mathrm{min}} \sum_{n=n_\mathrm{min}}^{n_\mathrm{max}-1} \left[ \frac{y^{(i)}_n - t^{(i)}_n}{\max\left\{\xi, t^{(i)}_n\right\}} \right]^2 \right\}^{1/2},
\end{equation}
for $i=1,2,3$, where $\xi\ge 1$ is a threshold and $n_\mathrm{min}$, $n_\mathrm{max}$ are such that
\begin{equation}
  \max\left\{ 0, n_\mathrm{ini} \right\} \le n_\mathrm{min} < n_\mathrm{max} \le \min\left\{N^\mathrm{(mod)} + n_\mathrm{ini}, N^\mathrm{(ref)}\right\}.
\end{equation}
In other words, $\mathsf{O}_{\vect{y},\vect{t}}$ is the sum of three functions, each of which considers one of the three reference quantities, separately. The model calibration is then performed by solving the following \textit{inverse problem}:
\par\noindent
\textit{Given} $\vect{p}^\mathrm{(fix)}$ \textit{and} $\vect{d}$\emph{, given the solution} $\vect{s}=\vect{g}\left(\vect{p}\right)$ \textit{to \eref{eq:DiscreteModelSynthetic}, determine} $\vect{y}\left(\vect{d},\vect{g}\left(\vect{p}\right),\vect{p}\right)$\textit{,} $\vect{t}$ \textit{and find} ${\vect{p}^{(\mathrm{cal})}}^\star$\textit{, such that}
\begin{equation}\label{eq:DefIP}
	\begin{array}{l}
	 \displaystyle {\vect{p}^{(\mathrm{cal})}}^\star = \arg\min_{\vect{p}^\mathrm{(cal)}\in\mathcal{P}^{(\mathrm{cal})}} \mathsf{O}_{\vect{y},\vect{t}}  \left( \vect{p}^\mathrm{(cal)} \right),\vspace{3pt}\\
	i.e.\vspace{3pt}\\
	\mathsf{O}_{\vect{y},\vect{t}}\left( {\vect{p}^{(\mathrm{cal})}}^\star \right) \le \mathsf{O}_{\vect{y},\vect{t}}\left( \vect{p}^\mathrm{(cal)} \right),\qquad\forall \vect{p}^\mathrm{(cal)}\:  :  \:
	\left( {\vect{p}^\mathrm{(fix)}}, {\vect{p}^\mathrm{(cal)}} \right) \in {\cal P}. 
	\end{array}
\end{equation}
In other words, the objective of model calibration is to find the parameter values which best fit the reference data in a given time interval, $n_\mathrm{min} \le n < n_\mathrm{max}$.

The threshold $\xi\in\mathbb{R}$, $\xi\ge 1$, plays a double role. First of all, it keeps positive the denominator of the quantity appearing in \eref{eq:ObjectiveFunctionExplicit}. Furthermore, it controls some characteristics of the objective function. For $\xi=1$, $\mathsf{O}_{\vect{y},\vect{t}}^{(i)}$ is nothing but the root-mean-squared relative difference between target ($t_n^{(i)}$) and modeled values ($y_n^{(i)}$) of the $i$-th component of $t_n$ and $y_n$. For larger values of $\xi$, relative errors corresponding to large values of $t_n^{(i)}$ will be dominant; from a practical point of view, this means that early time behavior is less relevant to the model fitting. In particular, if $\xi>\max\{t_n^{(i)},\,n_\mathrm{min}\le n < n_\mathrm{max}\}$, then $\mathsf{O}_{\vect{y},\vect{t}}^{(i)}$ reduces to the standard root-mean-squared error.

It is worth stressing that the explicit use of an interval $n_\mathrm{min} \le n < n_\mathrm{max}$ for the definition of $\vect{t}$, $\vect{y}$ and the objective function, although somehow cumbersome, is useful to assess changes in the physical parameters with time. Some examples of it will be shown in section \ref{sec:ModelCalibration}.

\subsection{Data and computer implementation for COVID-19}\label{sec:Data}
The application of the model introduced in section \ref{sec:Model} and of the model calibration introduced in section \ref{sec:InverseProblem} can be attempted thanks to publicly available data on COVID-19 pandemic. The application will be performed at national level, i.e., the considered population will be the whole population of some countries. For each country, the array $\vect{d}$ is populated with data coming from two basic sources.

Data on COVID-19 pandemic are available from the GitHub repository managed by the Johns Hopkins University \cite{JHU}. This is a collection of publicly available data from multiple sources, which are processed and delivered by the Johns Hopkins University Center for Systems Science and Engineering (JHU CSSE). Notice that the data are provided to the public strictly for educational and academic research purposes. The data are updated daily and the files used in this paper have been downloaded from the GitHub platform on May 2, 2020. The array $\vect{t}$ has been filled in by using those files.

\begin{figure}
  \centering
  \includegraphics[width=\textwidth]{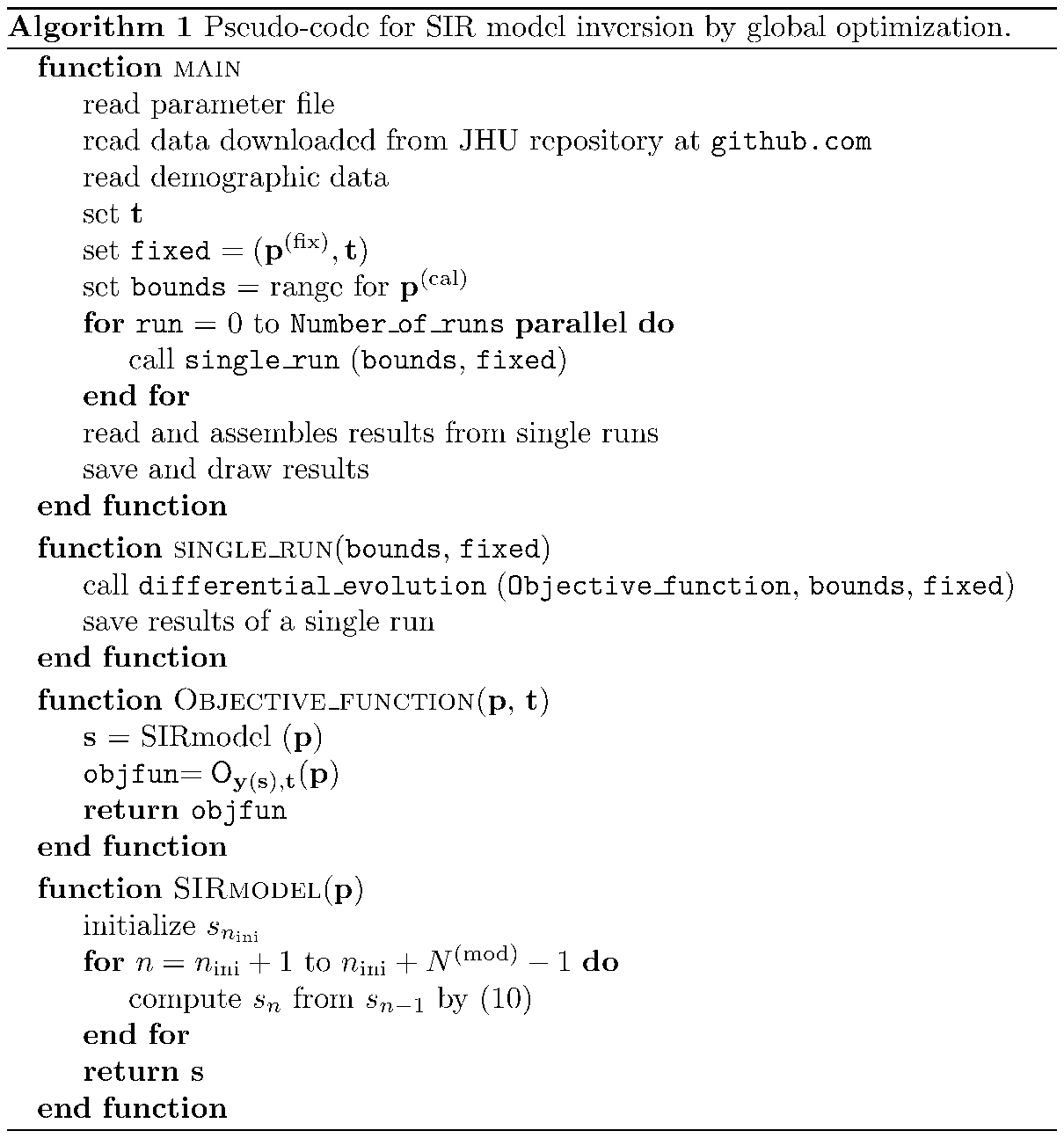}
\caption{Pseudo-code for SIR model inversion by global optimization.}\label{alg:pseudocode}
\end{figure}

Tailored codes have been developed under Python 3.7.6 to download data from the Github repository, perform the forward model introduced in subsection \ref{sec:Model} and calibrate the model by solving the corresponding inverse problem defined in subsection \ref{sec:InverseProblem}. The inversion is based on the functions of the \texttt{optimize} module from SciPy v1.4.1 and profit from multi-core execution through the standard \texttt{multiprocessing} package. The pseudo-code for inversion is given in 
Figure \ref{alg:pseudocode}. The optimization algorithms that have been tested are based on constrained minimization, so that some bounds on $\vect{p}^\mathrm{(cal)}$ should be prescribed. Best results have been obtained by global optimization with the function \texttt{differential\_evolution} \cite{StornPrice1997}. Since this function implements a stochastic algorithm, the pseudo-code of Figure \ref{alg:pseudocode} shows that several runs of the algorithms are executed in an easily parallelised loop.

\begin{figure}[htbp]
  \centering
  \includegraphics[width=\textwidth]{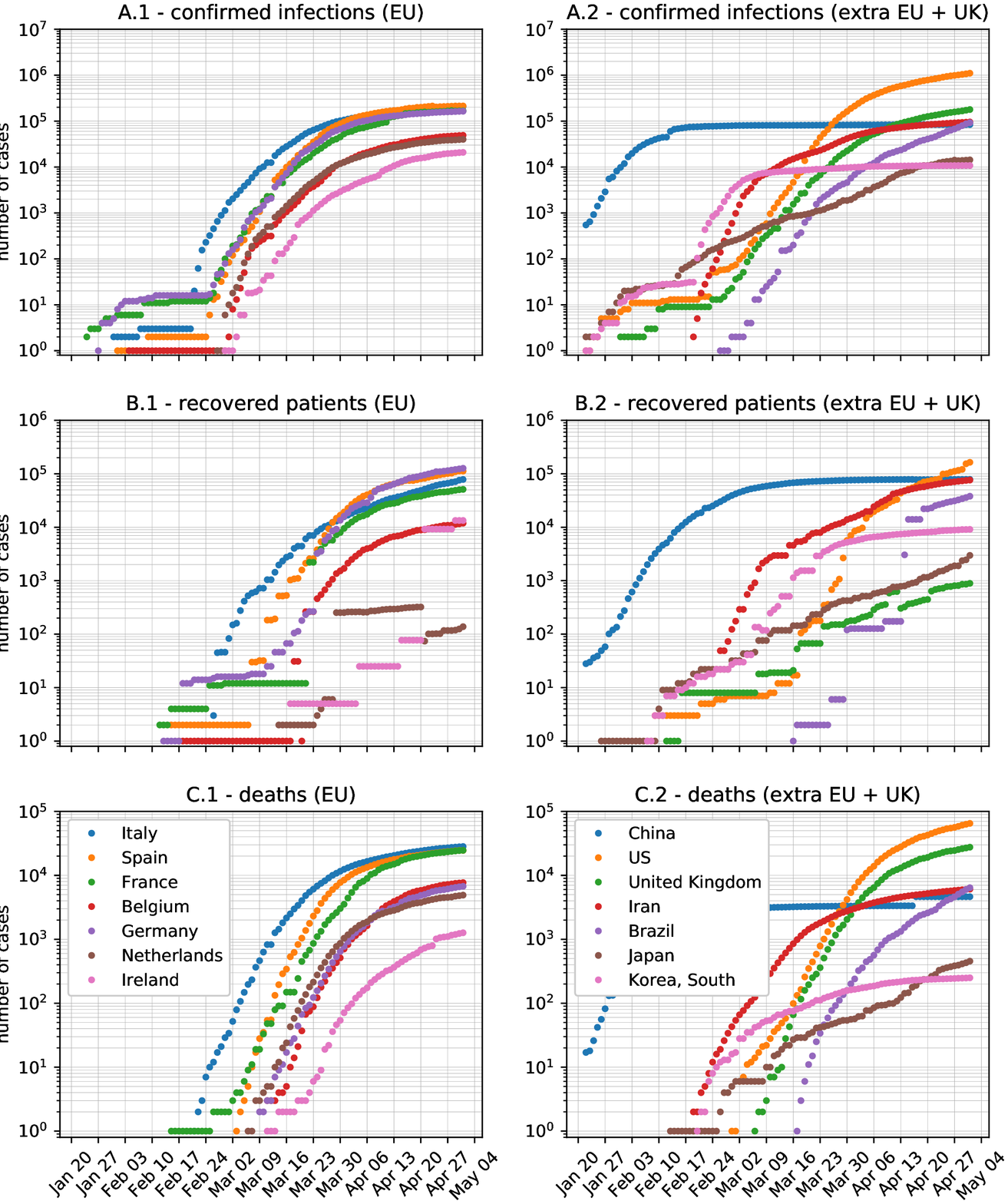}
  \caption{Data about COVID-19 pandemic in selected EU (left column) and extra EU + United Kingdom (right column) countries: A -- confirmed infections; B -- recovered patients; C -- deaths.}
  \label{fig:data}
\end{figure}

Figure \ref{fig:data} shows the trend of confirmed cases, recovered and deceased people for some countries, among those that have been considered as the most relevant for the analysis of COVID-19 pandemic not only by the scientific community, but also by mass media. These plots show different trends of the three curves describing the evolution in time of the confirmed, recovered and  dead cases among the various countries considered in this study.

Aside from China, for which the starting phase is not reported, since the virus diffusion started earlier than the first date for which data are available in the data set, the number of confirmed cases (plots A in Figure \ref{fig:data}) shows a first slow increase, followed by an exponential increase and possibly a slowdown after few weeks. It is highly questionable whether this behavior is related to the number of tests performed to confirm virus infection.

The most regular trends are clearly the ones describing the number of deceased people (plots C in Figure \ref{fig:data}), after about one week since the first reported case in each country considered in this study. Doubts about comprehensiveness of official data on deaths caused by coronavirus have been raised by several sources of information and by some commentators. Nevertheless, it seems safe to state that the number of deaths represents the time series with the smoothest variation and possibly the less affected by uncertainties in the data.

The second data source is the most updated version of the UN Demographic Yearbook \cite{UN2018DemographicYearBook}. Demographic data have been extracted from this volume. The values of population, birth and death rate of each country, for which the model has been tested, are included in $\vect{d}$. They are used to fix the values of $\beta$ and $\delta$ and to provide a first estimate of $P_\mathrm{ini}$.

Notice that the daily sampling rate of epidemiologic data induces to choose $\Delta t=1\,\mathrm{day}$. Moreover, the coefficients $\beta$ and $\delta$ are expressed on a daily basis, i.e., they are converted to the same measurement units as $\gamma$, $\phi$ and $\rho$, namely $\mathrm{day}^{-1}$ (see Definition \ref{definizione parametri}).

\section{Results}\label{sec:results}

\subsection{Model results}\label{sec:ModelResults}
First of all, the behavior of the model is shown with test case 1, which includes three model runs for which all the model parameters, but $\rho$, are kept fixed. The list of parameter values is given in Table \ref{tab:TestCase1}; the results of the model for a one-year-long simulation period are shown in Figure \ref{fig:TestCase1}. The general behavior shows an exponential increase in the number of infected persons (notice that the vertical axis is in logarithmic scale) followed by an exponential decrease but with a longer characteristic time. The number of deaths obviously decreases if $\rho$ increases and in particular, we have three different situations for the three runs: (a)~for the smallest value of $\rho$, the curve of susceptible persons dramatically decreases from some days before the peak of infections and reaches very small values after few weeks; (b)~for the intermediate value of $\rho$, the chosen values of model parameters yield a stationary conditions after about 8 months from the start of the epidemic for the number of susceptible and dead people, which reach almost the same value; (c)~for the highest value of $\rho$, the number of susceptible people decreases with time, but remains high. Notice that, for this test case, the reduction of the total population is limited, less than 10\%, and after one year almost all the living population is recovered. It is important to stress that this test case has the goal of showing how the model can predict different behavior and these results should not be considered as a forecast of the actual behavior of any real pandemic.

\begin{table}[htb]
\caption{Parameter values for test case 1.}
\label{tab:TestCase1}
\centering
\begin{tabular}{cccc}
\toprule
Parameter & run (a) & run (b) & run (c) \\
\midrule
$\beta$ & $2.46\cdot 10^{-5}\,\mathrm{day}^{-1}$ & {\small idem} & {\small idem} \\
$\delta$ & $3.01\cdot 10^{-5}\,\mathrm{day}^{-1}$ & {\small idem} & {\small idem} \\
$\gamma$ & $0.2\,\mathrm{day}^{-1}$ & {\small idem} & {\small idem} \\
$\rho$ & $0.01\,\mathrm{day}^{-1}$ & $0.05\,\mathrm{day}^{-1}$ & $0.1\,\mathrm{day}^{-1}$ \\
$\phi$ & $0.001\,\mathrm{day}^{-1}$ & {\small idem} & {\small idem} \\
$P_\mathrm{ini}$ & $10^9$ & {\small idem} & {\small idem} \\
\bottomrule
\end{tabular}
\end{table}

\begin{figure}[htbp]
  \centering
  \begin{subfigure}[b]{0.5\textwidth}
		\centering
	  \small (a) $P_\mathrm{ini} = 10^6$\\
    \includegraphics[width=\textwidth]{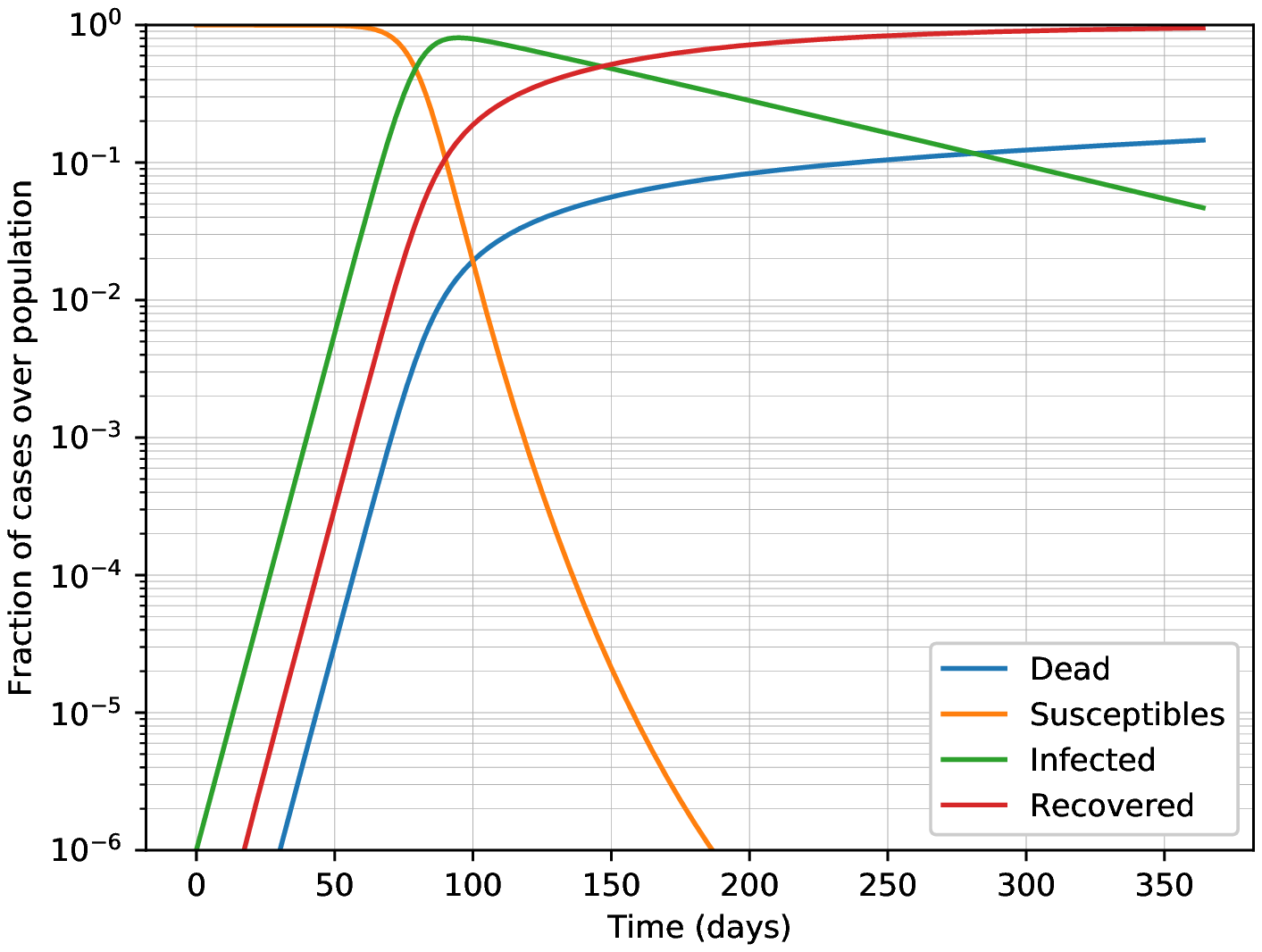}
  \end{subfigure}~
  \begin{subfigure}[b]{0.5\textwidth}
		\centering
	  \small (b) $P_\mathrm{ini} = 10^7$\\
	  \includegraphics[width=\textwidth]{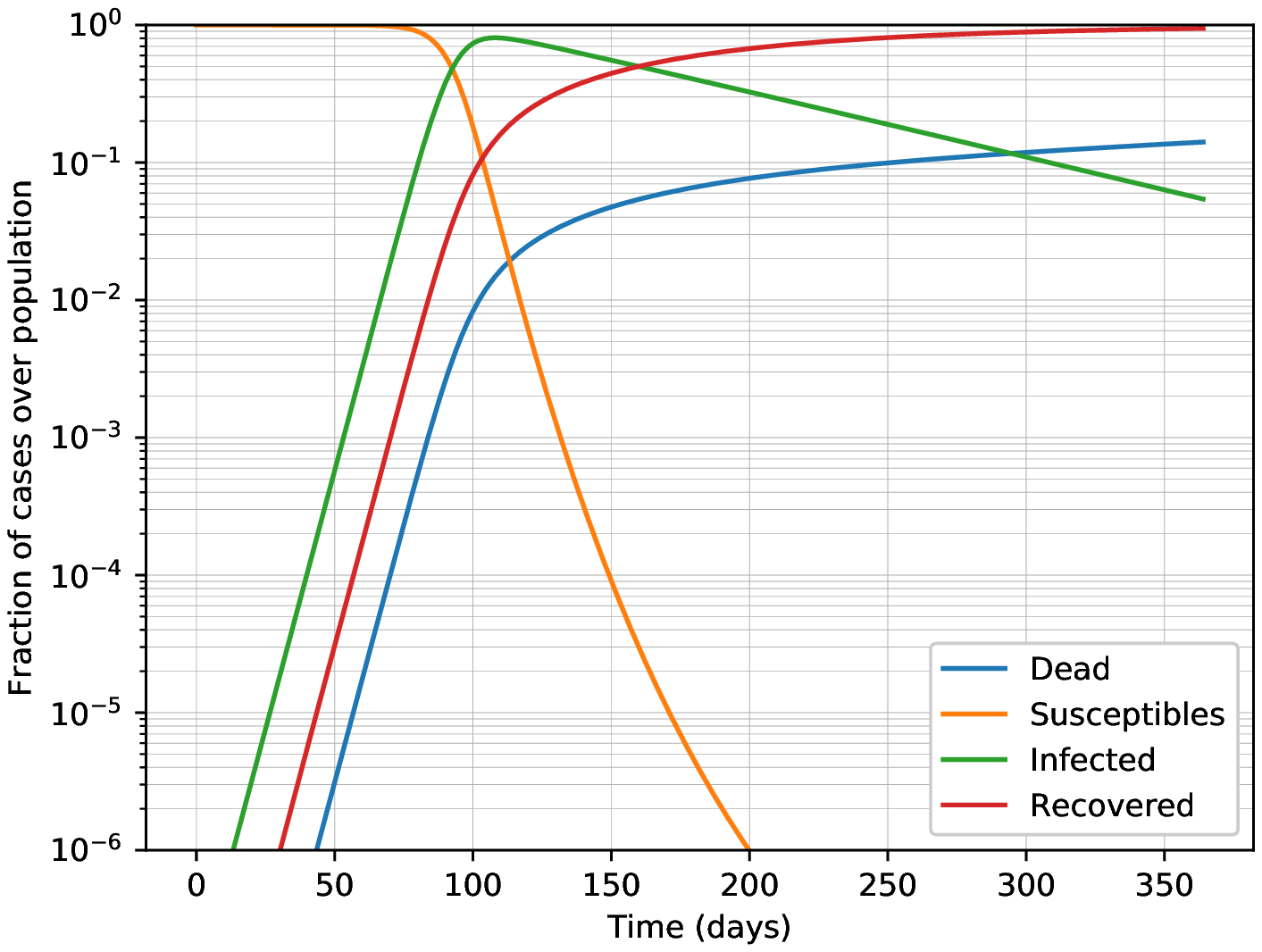}
  \end{subfigure}
		
  \begin{subfigure}[b]{0.5\textwidth}
		\centering
	  \small (c) $P_\mathrm{ini} = 10^8$\\
	  \includegraphics[width=\textwidth]{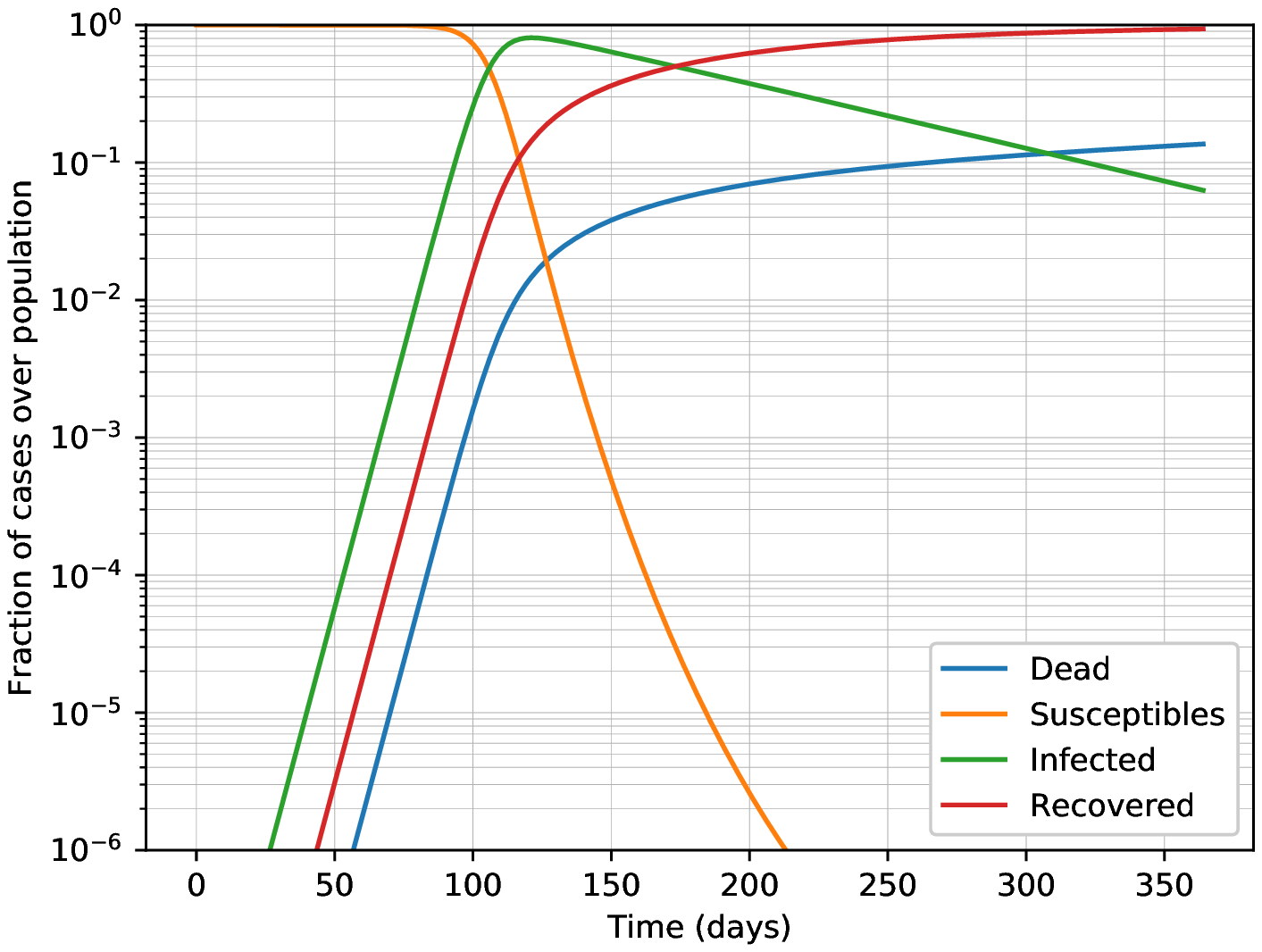}
  \end{subfigure}~
  \begin{subfigure}[b]{0.5\textwidth}
		\centering
	  \small (d) $P_\mathrm{ini} = 10^9$\\
	  \includegraphics[width=\textwidth]{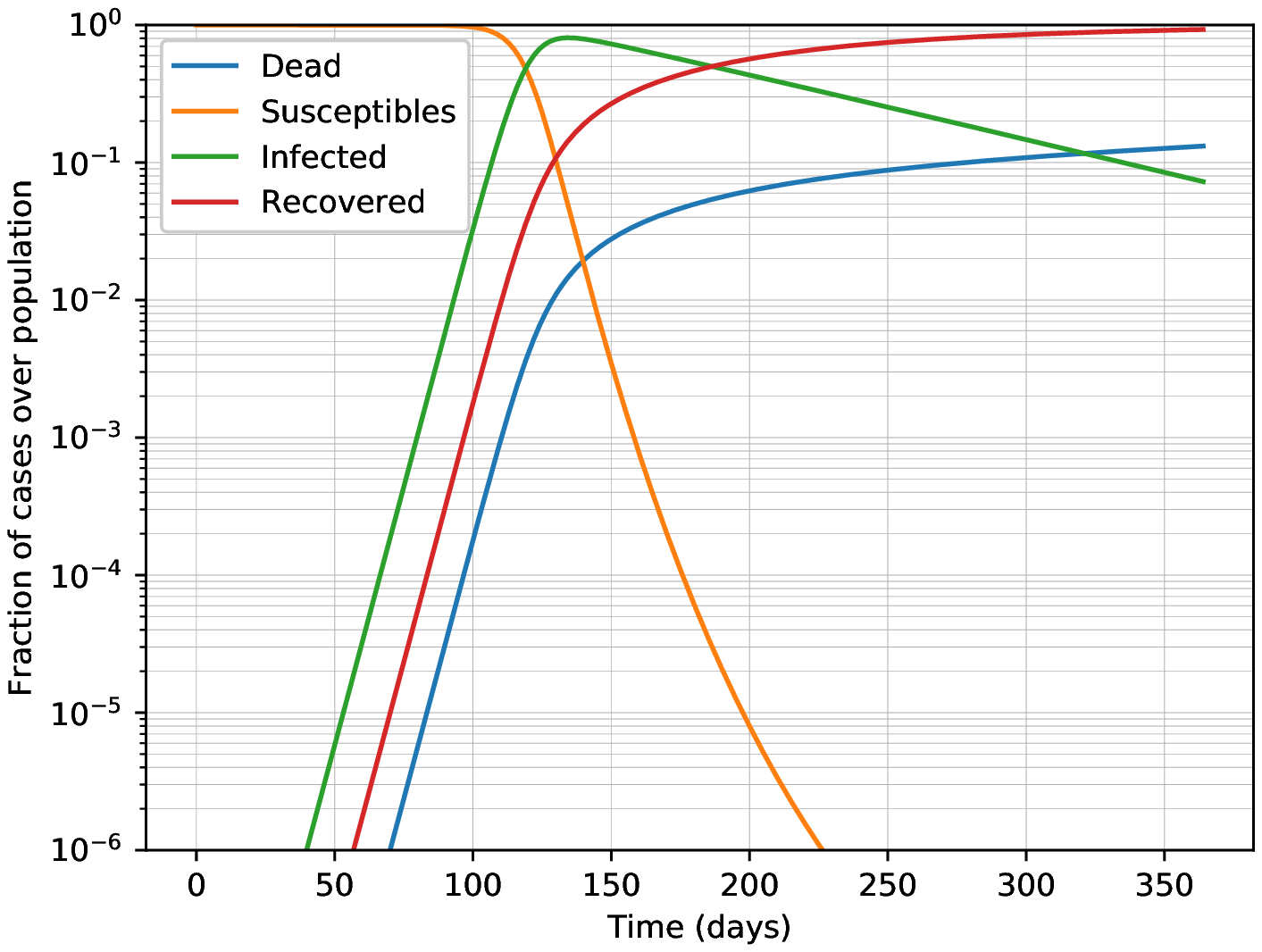}
  \end{subfigure}
  \caption{Model results for test case 2.}
  \label{fig:TestCase2_normalized}
\end{figure}

SIR models are often applied using the ratio of the number of individuals in each category with respect to the total population as state variables, namely $S/P$, $I/P$, $R/P$. Test case 1 showed that for three sets of model parameters, which differ only for the value of $\rho$, the total population has only a limited variation, so that approximating $P$ to a constant value could appear reasonable. Nevertheless, the term used to compute the infection rate is directly proportional to both $I$ and $S$ and inversely proportional to $P$ so that it introduces a non-linearity in the model. Therefore test case 2 is designed to assess the effect of $P_\mathrm{ini}$ on model results. To this goal, $P_\mathrm{ini}$ values span four orders of magnitude, from $10^6$ to $10^9$, whereas the other parameters are fixed at the values of run (a) of test case 1. The results are shown in Figure \ref{fig:TestCase2_normalized} as functions of the normalized quantities versus time. The values of each function at the end of the simulation period are very similar. The main differences are in the evolving phase, for which the response of a small population appears to be more rapid than that of a large population. Roughly speaking, the curves corresponding to high populations show a delay with respect to the curve for the smallest population of about 15 days per an increase in $P_\mathrm{ini}$ by an order of magnitude. This remark, if confirmed by runs with more reliable parameter sets, could have fundamental consequences in the design of early warning systems. In fact, the time at which a given threshold of cases over the total population is exceeded increases with the population size. 

\subsection{Model calibration}\label{sec:ModelCalibration}
\begin{figure}[htbp]
  \centering
	\begin{subfigure}[b]{0.5\textwidth}
	  \centering
	  \small (a) $\rho = 10^{-2}\,\mathrm{day}^{-1}$\\
    \includegraphics[width=\textwidth]{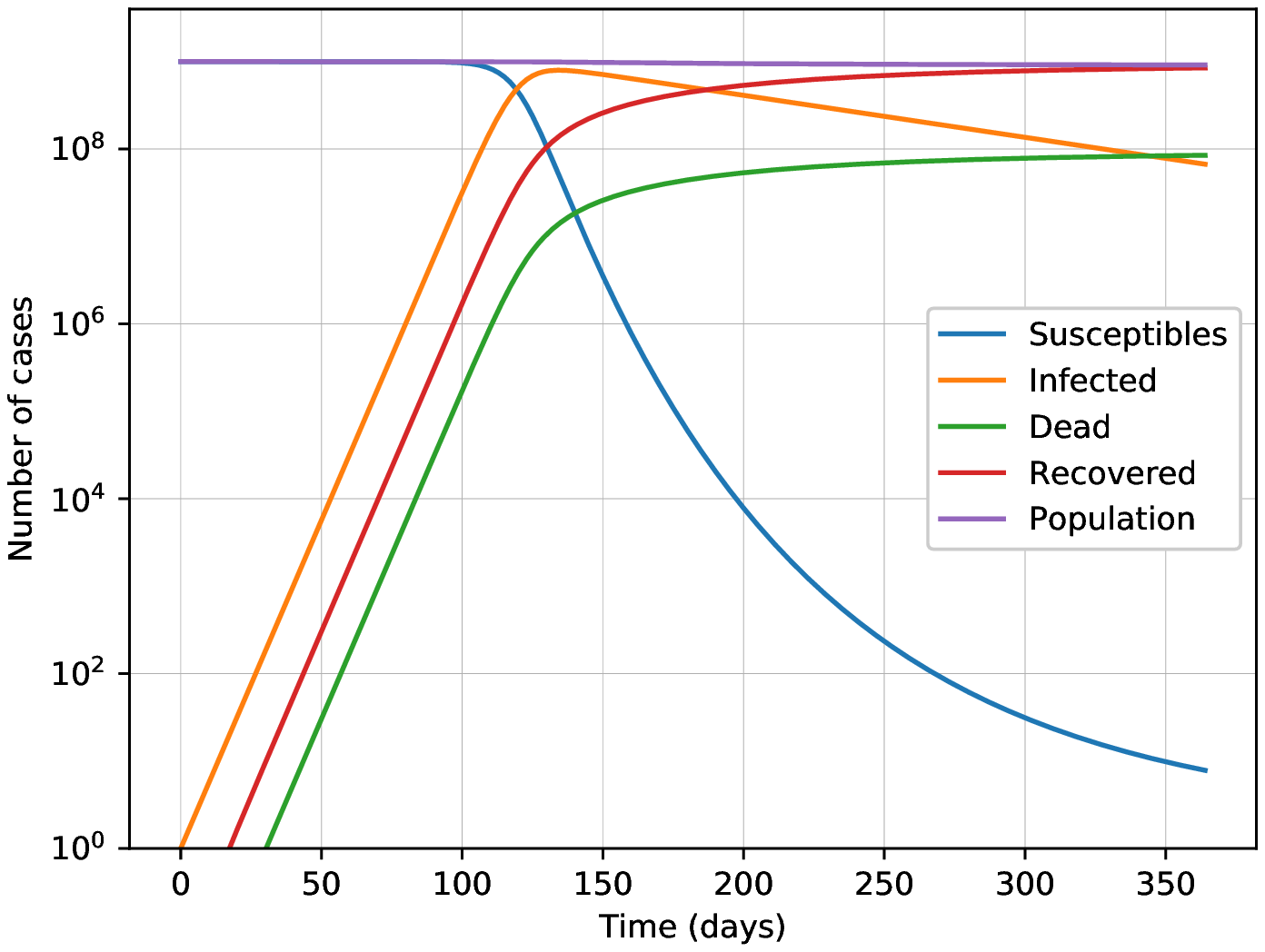}
	\end{subfigure}~
  \begin{subfigure}[b]{0.5\textwidth}
	  \centering
	  \small (b) $\rho = 5\cdot 10^{-2}\,\mathrm{day}^{-1}$\\
	  \includegraphics[width=\textwidth]{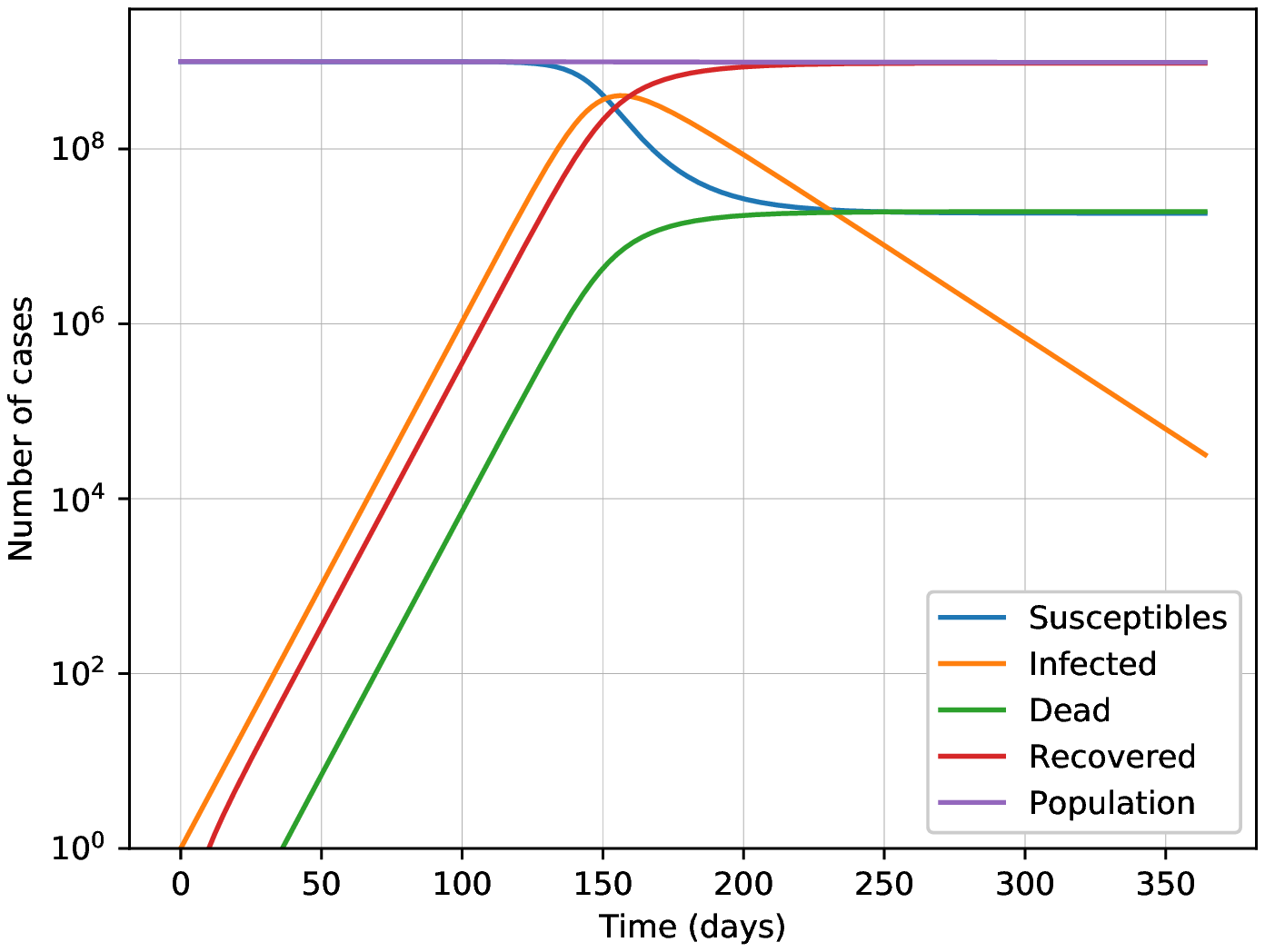}	
  \end{subfigure}

  \begin{subfigure}[b]{0.5\textwidth}
		\centering
	  \small (c) $\rho = 10^{-1}\,\mathrm{day}^{-1}$\\
	  \includegraphics[width=\textwidth]{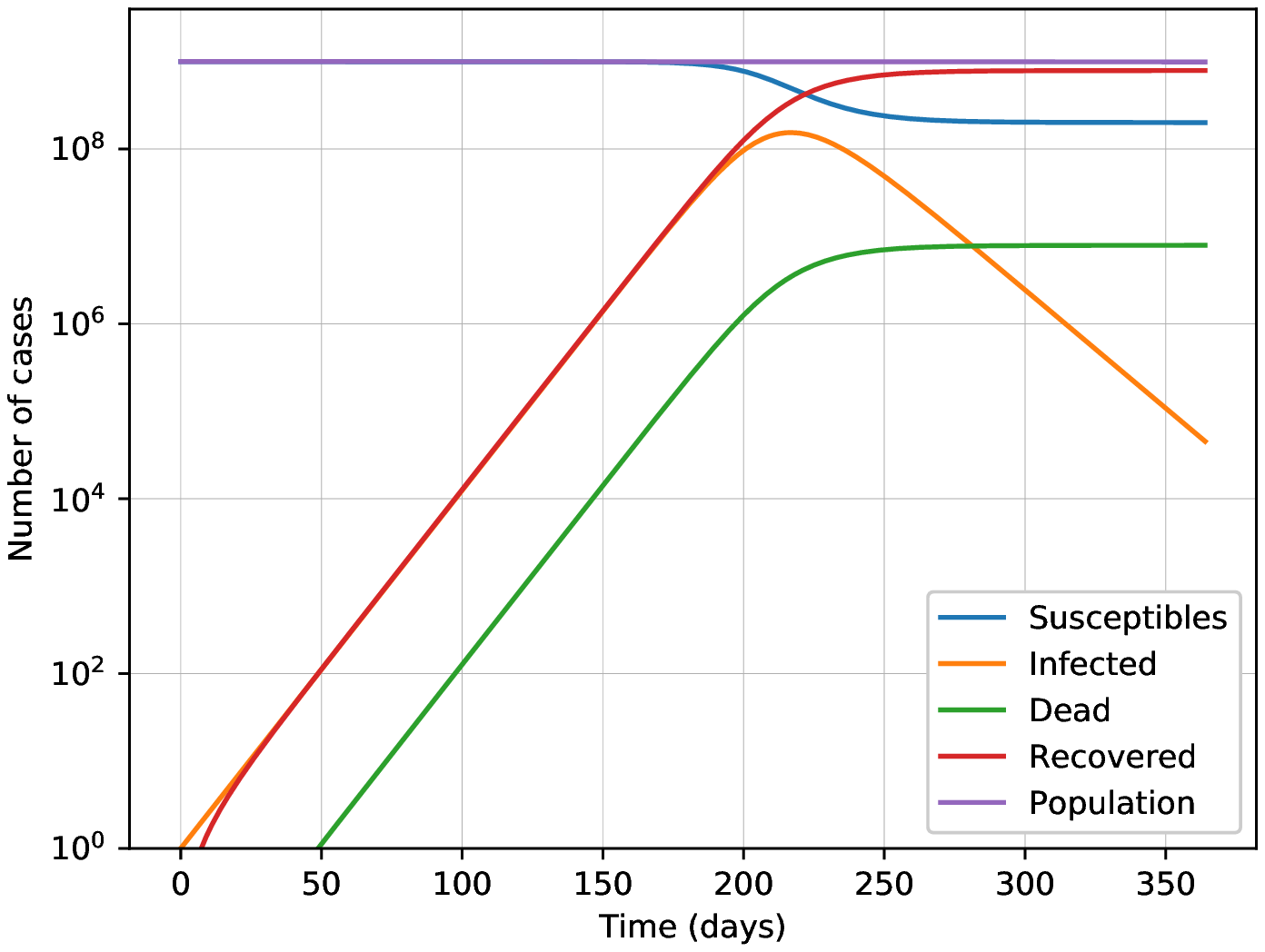}
  \end{subfigure}
  \caption{Model results for test case 1.}
  \label{fig:TestCase1}
\end{figure}

Model calibration for the COVID-19 pandemic by solution of the inverse problem is a very challenging problem. This is not surprising at all, because the comparison of the trends of the model time series (Figure \ref{fig:TestCase1}) with those observed from the reference data and drawn in Figure \ref{fig:data} shows that the SIR model can hardly reproduce the observed trend.

In particular, this paper is focused on the results obtained with data from Italy, but the same qualitative remarks hold also for the application to data from other countries.

\begin{figure}[htbp]
  \centering
  \includegraphics[height=0.33\textheight]{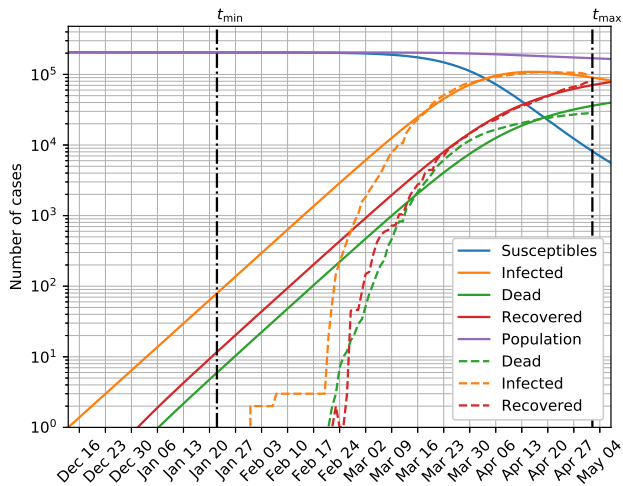}
  \caption{Comparison of reference (dashed lines) and modeled values (continuous lines) for Italy with the parameters obtained by solution of the inverse problem for test A (see Table \ref{tab:InversionTests}). The vertical dotted black lines delimit the time-frame of the data set used for model calibration, i.e., they correspond to $t_\mathrm{min}$ and $t_\mathrm{max}$.}
  \label{fig:TestA}
\end{figure}

The basic properties of the performed tests are listed in Table \ref{tab:InversionTests}. The comparison between reference and fitted time series for test A, which is to be considered as the ideal one, because all the data are used and the standard settings are applied, is shown in Figure \ref{fig:TestA}. The discrepancy between reference and modelled values in log scale is greater for the initial phase of the epidemic; the model does not reproduce the sharp reduction of the rate of increase of deaths which appears in the reference time series around mid March.

Notice that for tests B, C and D three subsets of data are used, corresponding to three non overlapping time intervals, each of which is 33-days-long. In particular, the first day for which data are available is January 22, 2020 and the data series used in this paper ends on May 2, 2020. Therefore, the data set considered for test B ends on February 24, 2020, whereas the data series used for test D starts on March 30, 2020. The goal of these three tests is to examine possible differences in the optimal values of the parameters and in the behavior of the inversion procedure, for successive temporal phases of the epidemic. For test E, $\xi=1$, so that each of the functions $\mathsf{O}_{\vect{y},\vect{t}}^{(i)}$ given by \eref{eq:ObjectiveFunctionExplicit} is nothing but the root-mean-squared relative difference between reference and modeled values of $I$, $R$ and $D$ for $i=1,2,3$, respectively. Test F is based on a subset of the data, in particular, for this test the number of dead patients only is fitted; the rationale behind this test is that $D^\mathrm{(ref)}$ should be less uncertain than the other data of $\vect{d}^\prime$. Finally, test G is an attempt to consider the hints raised by several authorities and researchers, suggesting that official numbers could be heavily underestimated. In this test, it is assumed that the number of infected and recovered persons be 10 times greater than those reported in official documents; analogously the number of deaths is assumed to be twice the official value.

\begin{table}[htb]
\caption{Inversion tests with data referred to Italy. The standard approach uses the settings described in subsection \ref{sec:InverseProblem} with $\xi = 10^6$ and the data described in subsection \ref{sec:Data}. Test G is based on the hypotheses that (i) the numbers of infected and recovered persons are ten times those reported by official fonts and (ii) the number of deaths is twice the official one.}
\label{tab:InversionTests}
\centering
\begin{tabular}{cccc}
\toprule
Test & $n_\mathrm{min}$ & $n_\mathrm{max}$ & Notes\\
\midrule
A & 0 & 101 & standard \\
B & 0 & 33 & standard \\
C & 34 & 67 & standard \\
D & 68 & 101 & standard \\
E & 0 & 101 & $\xi = 1$ \\
F & 0 & 101 & $\mathsf{O}^{(3)}_{\vect{y},\vect{t}} \left( \vect{p}^\mathrm{(cal)} \right)$ \\
G & 0 & 101 & modified data \\
\bottomrule
\end{tabular}
\end{table}

Minimization of the objective function $\mathsf{O}_{\vect{y},\vect{t}}$ was performed with different functions of the SciPy's module \texttt{optimize} which implements several methods to find a minimum, also by taking into account possible bounds on $\vect{p}^\mathrm{(cal)}$. The bounds have been assigned on the basis of preliminary gross estimates from available data and they are listed in Table \ref{tab:bounds}.

\begin{table}[htb]
\caption{Intervals of variation fixed for the parameters to be calibrated for inversion of data referred to Italy.}
\label{tab:bounds}
\centering
\begin{tabular}{lccccc}
\toprule
& $\gamma$ & $\rho$ & $\phi$ & $t_\mathrm{ini}$ & $P_\mathrm{ini}$ \\
\midrule
minimum & $10^{-4}\,\mathrm{day}^{-1}$ & $10^{-5}\,\mathrm{day}^{-1}$ & $10^{-6}\,\mathrm{day}^{-1}$ & $-60$ & $2\cdot 10^5$ \\
maximum & $1\,\mathrm{day}^{-1}$ & $0.1\,\mathrm{day}^{-1}$ & $0.1\,\mathrm{day}^{-1}$ & $20$ & $10^8$ \\
\bottomrule
\end{tabular}
\end{table}

Several runs have been conducted with a routine for local minimization and the best results were obtained with the L-BFGS-B method, which is a variation of the Broyden–Fletcher–Goldfarb–Shannon (BFGS) algorithm \cite{Fletcher2013} to reduce memory requirements and to handle simple constraints. The results of these runs are not presented here, for two basic motivations. That method is part of a wide family of algorithms which move towards the minimum by means of gradient-based searches.  However, it is not possible to compute analytically derivatives of $\mathsf{O}_{\vect{y},\vect{t}}$ with respect to $t_\mathrm{ini}$ and $P_\mathrm{ini}$, which are integer, and not real, variables. Therefore, that family of methods cannot be applied in a rigorous way. Nevertheless, the performed runs, possibly fixing the value of $t_\mathrm{ini}$, confirm the existence of multiple local minima for $\mathsf{O}_{\vect{y},\vect{t}}$.

Global minimization by application of \texttt{differential\_evolution} \cite{StornPrice1997}, even with the default settings, yielded good results, which are listed in Tables \ref{tab:results1} and \ref{tab:results2}. The mean value and its standard deviation of each parameter has been estimated after 10 runs of this stochastic algorithm, for which the random initializing seed introduces variations among the returned results. When looking at Table \ref{tab:results2}, it is important to recall that $t_\mathrm{ini}$ and $P_\mathrm{ini}$ are integer numbers, but in the table the averages and the relative standard errors are computed after 10 runs and this explains the float numbers notation.

Besides the optimal values of $\vect{p}^\mathrm{(cal)}$ listed in Tables \ref{tab:results1} and \ref{tab:results2}, it is important and useful to consider also some properties of the inversion procedure for each test; they are listed in Table \ref{tab:InversionParameters}.

\begin{table}[htb]
\caption{Results of model calibration by inversion of data referred to Italy for $\gamma$, $\rho$ and $\phi$. For the details about the performed tests see Table \ref{tab:InversionTests}.}
\label{tab:results1}
\centering
\begin{tabular}{cccc}
\toprule
Test& $\gamma$ (in $\mathrm{day}^{-1}$) & $\rho$ (in $\mathrm{day}^{-1}$) & $\phi$ (in $\mathrm{day}^{-1}$) \\
\midrule
A & $0.1381\pm 2\times 10^{-4}$ & $(1.761\pm 0.002)\times 10^{-2}$ & $(8.24\pm 0.02)\times 10^{-3}$ \\
B & $0.26\pm 0.05$ & $(3.27\pm 0.6)\times 10^{-3}$ & $(6\pm 1)\times 10^{-3}$ \\
C & $0.185\pm 4\times 10^{-3}$ & $(1.88\pm 0.01)\times 10^{-2}$ & $(1.52\pm 0.01)\times 10^{-2}$ \\
D & $0.12229\pm 1\times 10^{-5}$ & $(1.694\pm 0.001)\times 10^{-2}$ & $(7.929\pm 0.005)\times 10^{-3}$ \\
E & $0.17\pm 0.04$ & $(1.3\pm 0.2)\times 10^{-2}$ & $(1.2\pm 0.3)\times 10^{-2}$ \\
F & $0.1384\pm 2\times 10^{-4}$ & $(1.556\pm 0.002)\times 10^{-2}$ & $(1.450\pm 0.004)\times 10^{-3}$ \\
G & $0.28\pm 0.01$ & $(2.3\pm 0.2)\times 10^{-2}$ & $(1.2\pm 0.9)\times 10^{-2}$ \\
\bottomrule
\end{tabular}
\end{table}

\begin{table}[htb]
\caption{Results of model calibration by inversion of data referred to Italy for $t_\mathrm{ini}$ and $P_\mathrm{ini}$. For the details about the performed tests see Table \ref{tab:InversionTests}.}
\label{tab:results2}
\centering
\begin{tabular}{ccc}
\toprule
Test& $t_\mathrm{ini}$ & $P_\mathrm{ini}$ \\
\midrule
A & $(-42.8\pm 0.2)\,\mathrm{day}$ & $(2.111\pm 0.003)\times 10^{5}$ \\
B & $(7.6\pm 3)\,\mathrm{day}$ & $(5.44\pm 0.86)\times 10^{7}$ \\
C & $(-18.5\pm 2)\,\mathrm{day}$ & $(3.2\pm 1.2)\times 10^{5}$ \\
D & $-59\,\mathrm{day}$ & $(2.1940\pm 0.0084)\times 10^{5}$ \\
E & $(9.3\pm 2.8)\,\mathrm{day}$ & $(3.5\pm 1.1)\times 10^{7}$ \\
F & $(-56.4\pm 0.2)\,\mathrm{day}$ & $(2.042\pm 0.002)\times 10^{6}$ \\
G & $(-10.1\pm 5.8)\,\mathrm{day}$ & $(1.0\pm 0.6)\times 10^{7}$ \\
\bottomrule
\end{tabular}
\end{table}


\begin{table}[htb]
\caption{Properties of inversion of data referred to Italy; the values are based on 10 runs of the minimization algorithm for each test. For the details about the performed tests see Table \ref{tab:InversionTests}}
\label{tab:InversionParameters}
\centering
\begin{tabular}{cccc}
\toprule
Test & Minimum of $\mathsf{O}_{\vect{y},\vect{t}}$ & \begin{tabular}{@{}c@{}c@{}}Number of \\ iterations of \\ the algorithm\end{tabular} & \begin{tabular}{@{}c@{}c@{}}Maximum number \\ of evaluations of \\ $\mathsf{O}_{\vect{y},\vect{t}}$ for a single run\end{tabular}\\
\midrule
A & $(7.462\pm 0.001)\times 10^{-3}$ & $125 \pm 4$ & 11,316 \\
B & $(2.102\pm 0.03)\times 10^{-5}$ & $31 \pm 2$ & 3,312 \\
C & $(2.60\pm 1.8)\times 10^{-3}$ & $190 \pm 14$ & 19,338 \\
D & $(8.3164\pm 0.0008)\times 10^{-3}$ & $140 \pm 4$ & 11,802 \\
E & $2.35 \pm 0.09$ & $41 \pm 12$ & 8,859 \\
F & $(4.9\pm 3.9)\times 10^{-4}$ & $320 \pm 27$ & 31,566 \\
G & $(5.2032\pm 0.0013)\times 10^{-2}$ & $83 \pm 3$ & 7,788 \\
\bottomrule
\end{tabular}
\end{table}

Table \ref{tab:results1} shows that, apart from few tests, the optimal values of $\gamma$, $\rho$ and $\phi$ are relatively similar, sharing the same order of magnitude and the relationship $\gamma > \rho > \phi$ among different tests. These inequalities are violated by the results of test B and possibly of test E; test B refers to the very initial days of the epidemic, whereas test E refers to the use of relative errors in the computation of the objective function. Notice that using relative errors gives some more weight to the small values of the elements of $\vect{t}$, which are those recorded at the beginning of the epidemic. Therefore, these results are quite consistent. Notice also that tests B and E are the only tests for which $t_\mathrm{ini}>0$. In these tests, like in test G, the calibrated parameters display the highest coefficient of variation (the ratio between the standard deviation and the average); in other words, these are the tests for which the optimal values show more uncertainty.

Two facts should be mentioned about the results of test A shown in Table \ref{tab:results2}: first, $t_\mathrm{ini} < 0$, i.e., it seems that the infection started before the official appearance of the first confirmed case; second, $P_\mathrm{ini}$ is close to the lower bound chosen in Table \ref{tab:bounds}, so that the model predicts that the population which has been involved in the infection could be relatively small. These qualitative remarks are confirmed by most of the other tests. Notice, in particular, that even if one considers test E, which gives the highest average value of $P_\mathrm{ini}$ among different runs, the runs which yield the least values of $\mathsf{O}_{\vect{y},\vect{t}}$ give a value of $P_\mathrm{ini}$ close to $2\cdot 10^5$, as for test A.

\begin{figure}[htbp]
  \centering
  \begin{subfigure}[b]{0.5\textwidth}
		\centering
	  \small (a) optimal parameters\\
    \includegraphics[width=\textwidth]{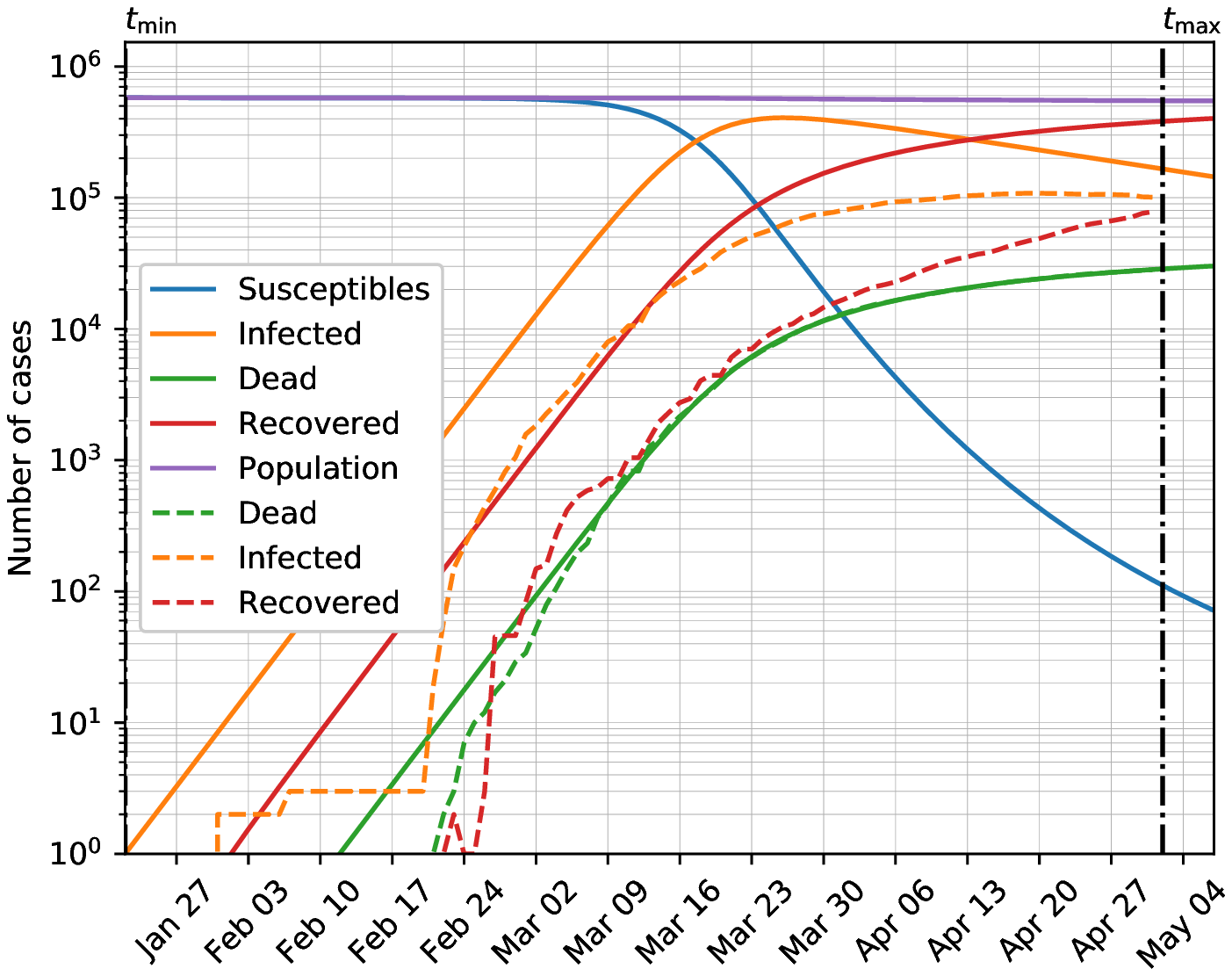}
  \end{subfigure}~
  \begin{subfigure}[b]{0.5\textwidth}
		\centering
	  \small (b) average parameters\\
	  \includegraphics[width=\textwidth]{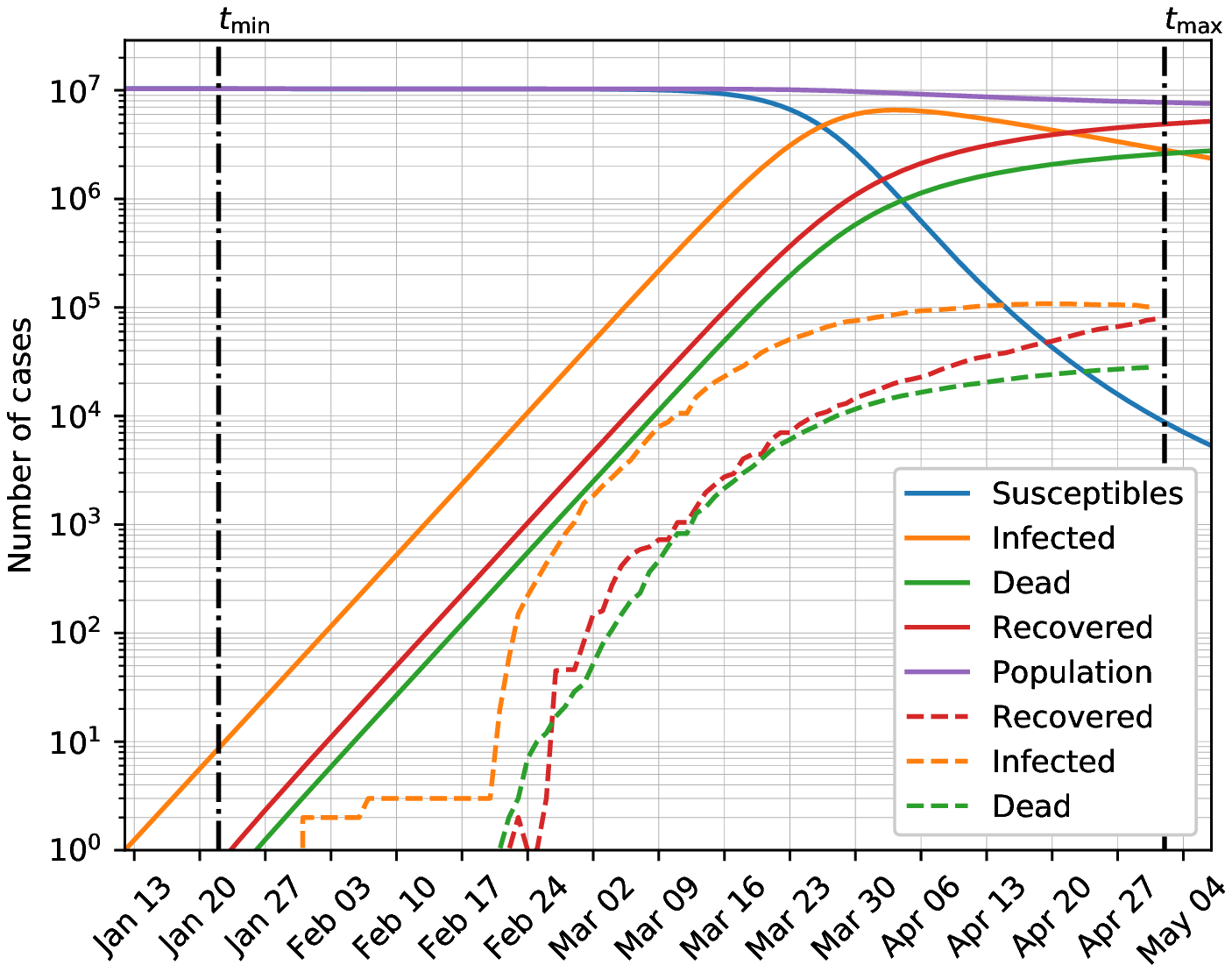}
  \end{subfigure}
  \caption{Comparison of reference (dashed lines) and modeled values (continuous lines) for Italy with the parameters obtained by solution of the inverse problem for test F (see Table \ref{tab:InversionTests}): (a)~optimal parameters corresponding to the global minimum; (b)~parameters averaged among the 10 inversion runs (Tables \ref{tab:results1} and \ref{tab:results2}). The vertical dotted black lines delimit the time-frame of the data set used for model calibration, i.e., they correspond to $t_\mathrm{min}$ and $t_\mathrm{max}$.}
  \label{fig:testF}
\end{figure}

Table \ref{tab:InversionParameters} shows that tests A, D and G are those for which the results of different runs are more consistent with each other. This is important, because it shows that the identification of ${\vect{p}^{(\mathrm{cal})}}^\star$ with the proposed approach appears to be robust for these tests. On the other hand, for the remaining tests, it is important to carefully check the outcomes of each single run. In fact, the initial seed could introduce some bias which cannot be overcome by the \texttt{differential\_evolution} routine with its default settings and the final result could yield a local minimum, instead of the global one. This is illustrated by the comparison in Figure \ref{fig:testF}, which shows the results of test F for the optimal parameters and those averaged among the 10 runs and listed in Tables \ref{tab:results1} and \ref{tab:results2}. This test was designed to fit the data on the deceased people and this is apparent from Figure \ref{fig:testF}(a); the fit seems extremely goo, in fact, the two green curves overlap almost perfectly for a large time interval. On the other hand, from Figure \ref{fig:testF}(b) it is evident that some of the inversion runs yielded parameters which do not permit to properly and satisfactorily reproduce the data.

From Table \ref{tab:InversionParameters}, it is also apparent that the objective function is computed a great number of times for each single run of the tests. This number strongly varies among the tests. Recall that each computation of $\mathsf{O}_{\vect{y},\vect{t}}$ requires a run of the model, so that the computational costs could become important. The tests discussed in this paper run on a PC with an \texttt{Intel core i7 9th Gen} processor; the execution time of a single run varied from 34 s for test B to 722 s for test F.

\section{Discussion}\label{sec:discussion}

\subsection{Remarks about the model}\label{sec:RemarksModel}
Some basic assumptions, on which the model developed in this work is funded, deserve to be recalled and discussed.

The developed model basically assumes ``homogeneity'' of the population. In other words, no distinction is done in terms of sex, age, economic wealth, health and wellness, working conditions, life style, home state, and any other, including genetic background. Also, the model assumes that the population under study is a closed system, thus disregarding variations induced by short-time, tourist or business travels, by intermediate-time mobility of students and workers, and by long-time effects of migrant fluxes.

The model is also independent of the climatic and environmental conditions, i.e., the processes considered by the model are assumed to be independent of the variability of weather conditions and environmental quality at any temporal and space scale. In particular, this means that neither sharp and rapid variations nor annual or seasonal cycling should affect these processes.

Epidemic models rarely consider birth and death rates, because the corresponding terms in the underlying equations are usually negligible. In this work, these terms have been kept, in order to facilitate this discussion. In particular, following the assumption of population homogeneity, it is assumed that infected pregnant women give birth to infected babies and that this occur at the same rate as for susceptible women.

With regard to infection rate, which is described by the term $\gamma IS/P$ in \eref{eq 1} and \eref{eq 2}, some remarks are in order. This term is computed by assuming that each infected individual has a given, constant number of contacts with other persons per unit time. Our model assumes that the number of persons who cannot be infected is $I+R$, so that the fraction of contacted persons who cannot be infected is given by $(I+R)/P$; on the other hand, the fraction of contacted individuals who can be infected is given by $S/P$. This is equivalent to assuming that recovered people become immune to the virus, an aspect which is not confirmed by the scientific community (see, e.g., \cite{Shietal2020}). Moreover, recovered people are assumed to be not infectious, which is the case if the response of their immune system is so fast that, once they come in contact with the virus again, the virus is destroyed by the immune system before it can be spread to susceptible persons. The $\gamma$ coefficient, due to the ``homogeneity'' assumption, is considered to be independent of the factors which have been recalled at the beginning of this subsection; in particular, working and living conditions could control the distance and the duration of contacts of infected - and therefore infectious - individuals with other persons.

The so-called recovery and fatality coefficients $\rho$ and $\phi$ are assumed to be constant. This is not based on the ``homogeneity'' assumption only. In fact, this implies that recovery and fatality are modeled as instantaneous processes, i.e., independent of the time passed since infection; moreover, no distinction is done among death or healing of infected people according to the strength of their symptoms and to the location where they are treated (home and hospitals non-intensive, or Intensive Care Units -- ICUs). The latter condition could be modeled by subdividing the class of infected people among sub-classes, e.g., asymptomatic, with light symptoms, admitted to hospital non-intensive care units, admitted to ICUs \cite{SIDARTHE,rouabah2020}.

\subsection{Remarks about model calibration by solution of the inverse problem}\label{sec:RemarksCalibration}
The results presented in section \ref{sec:ModelCalibration} show some of the classical, well known difficulties of non-linear least-squares inversion, in particular the dependence of the solution on the starting values, related to the existence of multiple local minima, and the flatness of the objective function around the local minima.

We obtained excellent results by applying the ``differential evolution'' algorithm \cite{StornPrice1997}. Obviously, different algorithms for global optimization could be tested, like, e.g., genetic algorithms \cite{Davis1991,rouabah2020}, particle swarm optimization \cite{KennedyEberhart1995}, simulated annealing \cite{kirkpatrick1983optimization}.

With reference to the specific example under study, it is necessary to stress some aspects, mostly related to the role of data in model calibration \cite{Giudici2001}.

First of all, the solutions obtained by means of a global optimization algorithm for high values of the threshold $\xi$ show that the optimal value of $P_\mathrm{ini}$ is smaller than the total Italian population. This parameter $P_\mathrm{ini}$ has been included in $\vect{p}^\mathrm{(cal)}$ with the objective of assessing the extension of the reference population. In other words, including $P_\mathrm{ini}$ among the parameters to be calibrated might provide a, possibly very rough, estimate of the width of the initial population whose evolution is represented by the model. In this particular instance, the results suggest that the reference initial population does not cover the whole country, but only a limited portion.

The latter remark seems to go in tandem with the well-known fact that in the countries most affected by COVID-19, the epidemic spread of the virus had mostly concentrated in specific areas: the province of Hubei, and above all the city of Wuhan, in China; the Lombardy region, and above all the provinces of Bergamo, Brescia, Lodi and Milan, in Italy; the city of New York in the USA; \^{I}le-de-France in France; Madrid and Catalunya in Spain; London in the UK.

Finally, it is quite difficult to assess the quality of data on the COVID-19 pandemic, but their uncertainty is expected to be very high. For instance, the correct number of infected people ``remains unknown because asymptomatic cases or patients with very mild symptoms might not be tested and will not be identified'', as recognized, e.g., by \cite{Baud2020}. In an interview published on March 23rd 2020 by the Italian newspaper ``La Repubblica'', Angelo Borrelli, head of Dipartimento della Protezione civile (national civil protection department) stated that a ratio of one certified case out of every 10 total cases is credible. Furthermore, different criteria have been adopted by different countries and institutions to define the various categories of infected, recovered and deceased people by or with COVID-19. This fact has been widely recognized as a cause of uncertainty in the collected data. Finally, censorship on COVID-19 pandemics is reported by journalists and organizations in some of the countries affected by the pandemic.

As a consequence, the use of official data to perform reliable estimates is questionable. In principle, stochastic approaches, e.g., the Bayesian framework \cite{calvetti2020}, could be very helpful to handle discrepancies between model predictions and reference values. Unfortunately, in this case the systematic and random errors could be so high as to make it very difficult to handle them even in a stochastic framework.

\section{Conclusions}\label{sec:conclusions}
The modeling tests conducted within this work lead to a series of remarks, which are summarized in this conclusive section, together with some future perspectives.

Starting from some remarks about modeling aspects, the limitations of classical SIR models have been recalled. These should be always carefully considered especially for applications and when these models are used as engines of decision support systems.

The main limitation of our model is related to the ``homogeneity'' assumption, accompanied by the steadiness of the recovery and fatality coefficients.

The latter aspect could be handled, for instance, by introducing functions ($\tilde\phi$, $\tilde\rho$) of elapsed time since infection. Such functions should enter in a deconvolution product involving the number of persons who have been infected at a given time and are still infected, i.e., are not yet recovered or passed away. With this approach, $\phi I$ and $\rho I$ in \eref{eq 2} to \eref{eq 4} could be replaced by
\begin{equation}\label{eq:Modification}
  \begin{array}{c}
	\displaystyle \int_0^{\tau_\mathrm{max}} \tilde\phi(\tau) \tilde I(t - \tau)\,\mathrm{d}\tau \quad \mbox{and} \quad \int_0^{\tau_\mathrm{max}} \tilde\rho(\tau) \tilde I(t - \tau)\,\mathrm{d}\tau,\vspace{6pt}\\
	\mbox{where}\quad \displaystyle \tilde I(t - \tau) = \frac{\mathrm{d}I}{\mathrm{d}t} (t - \tau)\exp{ \left\{ - \int_0^\tau \left[ \tilde\rho(\tau') + \tilde\phi(\tau') \right]\,\mathrm{d}\tau' - \delta \tau \right\} }
	\end{array}
\end{equation}
and $\delta$ is the death rate introduced in definition \ref{definizione parametri}. Notice that the fatality coefficient, $\phi$, accounts for the deaths related to the pandemic, i.e., it represents the increase in the death rate due to the pandemic. The normal death rate is considered through $\delta$.

The assumption of ``homogeneity'' could be relaxed by considering ``distributed'' models, similar to those applied for transport phenomena, e.g., for diffusion of contaminants in the environment. Those models can account for ``diffusive'' spread and for ``advective'' transport. However, the required parametrization is often much finer than the one for lumped models, so that the number of parameters to be calibrated strongly increases, and therefore in absence of good quality data it could be difficult to perform a reliable calibration and validation of the model for a practical application.

Promising classes of models are stochastic models \cite{Isham1993}, either under a Monte Carlo framework or by using assimilation techniques, e.g., the Ensemble Kalman Filter (EnKF, see, e.g., \cite{Evensen2003}). In principle, Monte Carlo models might be adapted in a relatively easy way to account for several phenomena and also to consider the role of some aspects (e.g., sex, age, health and wellness, etc.) on the probability of infection, recovery and decease. On the other hand, EnKF could provide a firm theoretical framework to improve model predictions by means of uncertain data.

With regard to the specific application of our model to COVID-19 epidemic, although it could be improvident to draw quantitative conclusions, it is nevertheless qualitatively confirmed that infection started quite earlier than the certain appearance of the first episodes of infection. The results of model inversion also suggest that the calibrated model could be reliable for a portion of the whole population. Somehow, the model itself, through its calibration, seems to suggest the width of the population for which its approximations could be valid.

By merging inversion results with the analysis of the continuous model, some relevant remarks can be given for the conditions at the apex of infection, i.e., when $I$ reaches its maximum value. At that time,
\begin{equation}
  \frac{\mathrm{d}I}{\mathrm{d}t}=0\quad\Rightarrow\quad \beta + \gamma \frac{S}{P} - \delta -\phi -\rho = 0,
\end{equation}
and, after simple algebraic manipulations,
\begin{equation}
  I + R = \frac{\gamma - \alpha}{\gamma} P.
\end{equation}
The calibration results listed in Table \ref{tab:results1} show that $\alpha$ is about one order of magnitude smaller than $\gamma$. In particular for the calibration tests performed in this study (Table \ref{tab:results1}) $(\gamma - \alpha)\cdot\gamma^{-1}$ assumes a relatively high value, close to 0.8. In other words, if the values of $\gamma$, $\phi$ and $\rho$ obtained from model calibration could be considered as reliable, at least as order of magnitude, at the pandemic peak a large fraction of the population would have already been infected, and possibly recovered.

Last, but not least for its practical importance, this paper has the ambition to provide further evidence about the great care that has to be given to the quality of pandemic data, when used to calibrate or validate epidemic models. In fact, poor quality data might yield unrealistic parameter values and, therefore, unreliable model predictions.

\section*{Acknowledgments}
This work presents results of a purely curiosity-driven research, which has received support only through the standard working facilities of the authors' institutions. This research did not receive any specific grant from funding agencies in the public, commercial, or not-for-profit sectors.

The data on COVID-19 epidemic have been downloaded from the following URL: \noindent\texttt{from https://github.com/CSSEGISandData/COVID-19}.

\section*{Authors' contributions}
MG designed the work and forward and inverse modeling, wrote the first draft of the manuscript, developed the first version of the computer codes and performed calculations. AC improved the implementation in the computer codes. RG revised the analytical and formal mathematical aspects of the work. All authors equally contributed to the discussion of the results and to the revision of the final manuscript.

\bibliographystyle{amsplain}
\bibliography{Articolo_arXiv_COVID-19}     

%
%

\affiliations
\address{Mauro Giudici}{Universit\`a degli Studi di Milano, Dipartimento di Scienze della Terra ``A.Desio'', Milano, Italy}{mauro.giudici@unimi.it}

\address{Alessandro Comunian}{Universit\`a degli Studi di Milano, Dipartimento di Scienze della Terra ``A.Desio'', Milano, Italy}{alessandro.comunian@unimi.it}

\address{Romina Gaburro}{University of Limerick, Department of Mathematics and Statistics, Health Research Institute (HRI), Limerick, Ireland}{romina.gaburro@ul.ie}


\end{document}